\documentclass[12pt]{article}

\let\plural=\relax

\def\modifymargins#1#2{
\newdimen\addtoh
\newdimen\addtow
\addtoh=#1
\addtow=#2

\advance\topmargin by -\addtoh
\multiply\addtoh by 2
\advance\textheight by \addtoh

\advance\oddsidemargin by -\addtow
\advance\evensidemargin by -\addtow
\multiply\addtow by 2
\advance\textwidth by \addtow
}

\def\filename{\jobname.tex}

\let\simpleinput=\input

\def\input#1{%
\expandafter\let\csname prev#1\endcsname=\filename
\edef\filename{#1}
\simpleinput #1\relax
\edef\filename{
\expandafter\csname prev\filename\endcsname}}

\let\simplebibliography=\bibliography

\expandafter\ifx\csname bibliography\endcsname\relax
\else
\def\bibliography#1{\def\filename{\jobname.bbl}\simplebibliography{#1}}
\fi

\expandafter\ifx\csname AtEndDocument\endcsname\relax
\else
\AtEndDocument{\let\input=\simpleinput}
\fi

\def\space{ }
\def\pdfposition{[ @thispage /XYZ @xpos @ypos ]}

\let\preveverypar=\everypar
\newtoks\everypar
\preveverypar={
\the\everypar
}

\let\prevlabel=\label
\def\label#1{%
\prevlabel{#1}%
}

\makeatletter

\long\def\state#1#2#3#4#5{%
\def\statelabel{\label{#1}}
\def\st@atehere{#3#4}
\ifx\st@atehere\empty\expandafter\def\csname st@atementempty#1\endcsname{Y}\else#3\fi#4%
\expandafter\ifx\csname stateonce\endcsname\relax\relax\else #5\fi
\ifx#2\relax\else
\expandafter\def\csname st@atementlabel#1\endcsname{#2}
\fi
\expandafter\def\csname st@atement#1\endcsname{#4#5}
}

\newcount\restatec@unter

\def\restate#1{
\expandafter\ifx\csname stateonce\endcsname\relax
\expandafter\let\expandafter\what\csname st@atementlabel#1\endcsname
\expandafter\ifx\what\relax
\csname st@atement#1\endcsname
\else
\def\rest@teempty{Y}
\expandafter\ifx\csname st@atementempty#1\endcsname\rest@teempty
\def\statelabel{\label{#1}}
\csname st@atement#1\endcsname
\else
\expandafter\let\expandafter\counter\csname c@\what\endcsname
\restatec@unter=\counter
\counter=\expandafter\ifx\csname r@#1\endcsname\relax0\else\@kernel@ref{#1}\fi
\advance\counter by -1
\let\statelabel=\relax
\csname st@atement#1\endcsname
\counter=\restatec@unter
\fi
\fi
\fi
}

\makeatother

\expandafter\ifx\csname stateno\endcsname\relax
\else
\def\state#1#2{\def\statelabel{\label{#1}}}
\fi

\newenvironment{hfigure}{\refstepcounter{figure}\begin{center}}{\end{center}}
\def\hcaption#1{\par Figure \thefigure: #1}

\def\v{\begingroup\obeyspaces\obeylines\hsize 20cm\tt}
\def\vv{\endgroup}

\def\skipproof#1{\def\nope{\iffalse}\def\go{\iftrue}%
\ifx#1y[proof commented out]\let\next=\nope\else\let\next=\go\fi\next}

\def\draft{\iftrue\begingroup\par\noindent\em=============[draft]=============\par}
\def\enddraft{\par\noindent=============[/draft]=============\par\endgroup\fi}
\def\draft{\iffalse}
\let\enddraft=\fi

\def\rev{\mathrm{rev}}		
\def\lex{\mathrm{lex}}		
\def\nat{\mathrm{nat}}		
\def\res{\mathrm{res}}		
\def\vrad{\mathrm{vrad}}	
\def\sev{\mathrm{sev}}		
\def\msev{\mathrm{msev}}	
\def\dsev{\mathrm{dsev}}	
\def\full{\mathrm{full}}	

\makeatletter
\DeclareRobustCommand\imin{\mathop{\operator@font imin}}
\DeclareRobustCommand\imax{\mathop{\operator@font imax}}
\makeatother

\long\def\nop#1{}

\def\np{{\rm NP}}
\def\conp{{\rm coNP}}
\def\D#1{\mbox{$\Delta^p_{#1}$}}
\def\dpi{${\rm D}^p$}

\def\NOT{\mbox{not }}
\def\AND{\mbox{ and }}
\def\OR{\mbox{ or }}
\def\IFF{\mbox{ iff }}

\def\true{{\sf true}}
\def\false{{\sf false}}
\def\proof{\noindent {\sl Proof.\ \ }}
\def\qed{\hfill{\boxit{}}
  \ifdim\lastskip<\medskipamount \removelastskip\penalty55\medskip\fi}
\long\def\boxit#1{\vbox{\hrule\hbox{\vrule\kern3pt
                  \vbox{\kern3pt#1\kern3pt}\kern3pt\vrule}\hrule}}

\def\color[#1]#2{}
\def\possnewtheorem#1#2{
\expandafter\ifx\csname #1\endcsname\relax
\newtheorem{#1}{#2}
\fi
}

\possnewtheorem{theorem}{Theorem}
\possnewtheorem{corollary}{Corollary}
\possnewtheorem{lemma}{Lemma}
\possnewtheorem{definition}{Definition}
\possnewtheorem{algorithm}{Algorithm}
\possnewtheorem{counterexample}{Counterexample}

\expandafter\ifx\csname ttytty\endcsname\relax
\else
\modifymargins{0cm}{-10pt}
\fi

\title{Forgetting in short and heterogeneous sequences of belief revisions}
\author{Paolo Liberatore%
\thanks{
DIAG, Sapienza University of Rome,
{\tt liberato@diag.uniroma1.it}
}
}

\begin{document}

\maketitle

\draft

FARE: separator, newpage, draft, FARE

formule troppo lunghe

\enddraft

\draft

{\bf PER RISPONDERE, se rilevante}

nelle dimostrazioni di hardness e' preferibile che la revisione ridondante sia
la prima e che l'ordine sia piatto; non sono gli unici casi possibili, ma e'
piu' probabile che si dimentichi una revisione vecchia che nuova, e l'ordine
piatto richiede il caso comune di completa ignoranza iniziale; la prima e' una
cosa molto plausibile e la seconda puo' capitare; non e' detto che sia cosi', e
in particolare non per l'ordine piatto, ma sono due casi di interesse, quindi
conviene che la hardness sia dimostrata in questo caso; naturalmente, come e'
ovvio e chiunque capirebbe anche da solo, la hardness del caso generale e'
automaticamente dimostrata da questo; non avrebbe neanche senso chiedere una
dimostrazione nel caso generale

e' detto alla fine delle definizioni, ma non viene dato molto risalto

\enddraft

\draft

{\bf SVILUPPI ULTERIORI}

FARE: anche per gli OCF esiste una rappresentazione sintattica, un insieme di
(formula,numero) invece di (mondo,numero); il punto non e' tanto la esistenza
di una rappresentazione sintattica, ma quanto la sua compattezza, che poi e'
quello che indica se e' realistica

questo articolo riassume i meccanismi: paper-5.pdf
InfOCF-Lib: A Java Library for OCF-based Conditional Inference,
Steven Kutsch

"Several semantics have been proposed for defining nonmonotonic inference
relations over sets of such rules. Examples are Lewis' system of spheres [17],
conditional objects evaluated using Boolean intervals [9], or possibility
measures [8, 10]."

"Here, we will consider Spohn's ranking functions [20] that assign degrees of
disbelief to propositional interpretations. C-Representations [13, 14] are a
subset of all ranking functions for a knowledge base, which can conveniently be
calculated by solving a constraint satisfaction problem dependent on the
knowledge base. [...] the tool InfOCF [...] allows the user to load knowledge
bases, calculate admissible ranking functions (in particular c-representations
and system Z [19]), and perform inference using these sets of ranking
functions. System P entailment [1] was also implemented."

"several new theoretical approaches have been proposed such as different
notions of minimal c-representations [2] and inference using sets of ranking
functions [5, 7]."

FARE: altri sistemi pratici, rivedere cosa fanno di preciso:

1812.08313.pdf
Iterated belief revision under resource constraints: logic as geometry
Dan P. Guralnik, Daniel E. Koditschek

aaai.v33i01.33013076.pdf
Iterated Belief Base Revision: A Dynamic Epistemic Logic Approach
Marlo Souza, Alvaro Moreira, Renata Vieira
sembra includere lexicographic o essere equivalente

sistemi che estendono lexicographic:

- priority graph
  aaai.v33i01.33013076.pdf
  Iterated Belief Base Revision: A Dynamic Epistemic Logic Approach
  Marlo Souza, Alvaro Moreira, Renata Vieira
  2019

- combinazioni lessicografiche di ordinamenti generici
  document (11).pdf
  Operators and Laws for Combining Preference Relations
  Hajnal Andreka, Mark Ryan, Pierre-Yves Schobbens
  2002

FARE: vedere altro in iterated-revision.txt

\enddraft

\sloppy

\draft
\section{Synopsis}

\subsection{introduction.tex}

\begin{itemize}

	\item why redundancy is interesting:
		forgetting happens, does it cause information loss?

	\item results, open in previous article:

	\begin{itemize}

		\item heterogeneous sequences:
			strictly harder than homogeneous lexicographic

			same complexity of inference, why the difference?
			different complexity of comparison

		\item short Horn lexicographic sequences:
			polynomial for sequences of two

			side results: necessary and sufficient condition,
			polynomial algorithm for equivalence horn - not horn

	\end{itemize}

\end{itemize}

\subsection{definitions.tex}

	\begin{itemize}

		\item doxastic state as sequence of equivalence classes

		\item notation simplifications:
			omitted conjunctions,
			formulae as sets of models

		\item belief revision: lexicographic, natural, etc.

		\item redundancy of a revision

	\end{itemize}

\subsection{mixed.tex}

	\begin{itemize}

		\item redundancy is \dpi-hard for natural + severe

		\item \np-hard for deep severe + natural from flat state

	\end{itemize}

\subsection{comparison.tex}

	\begin{itemize}

		\item why same for inference but different in redundancy?

		\item base problem is comparison $I \leq J$, not inference

		\item revisions with three properties are \dpi-hard;
			natural, severe, restrained have them

		\item full meet is \np-hard

		\item moderate severe is \conp-hard

		\item deep severe is \np-hard

		\item lexicographic and very radical are polynomial

	\end{itemize}

\subsection{condition.tex}

	\begin{itemize}

		\item redundancy of $S_1$ in $\emptyset [S_1,\ldots,S_m]$
			as $S_1$ equivalent to the disjunction of some
			Q-combinations $B_2 \equiv S_2 \ldots B_m \equiv S_m$

		\item subcase for two formulae

		\item some necessary only conditions:
			prepending or appending revisions

	\end{itemize}

\subsection{horn.tex}

  \subsection{horn-two.tex}

	\begin{itemize}

		\item two lexicographic Horn formulae: polynomial

		\item algorithm for equivalence horn with negation of horn

	\end{itemize}

  \subsection{horn-three.tex}

	\begin{itemize}

		\item three lexicographic Horn formulae: open

	\end{itemize}

\subsection{conclusions.tex}

	\begin{itemize}

		\item summary of results

		\item open problems

	\end{itemize}

\enddraft

\begin{abstract}

Forgetting a specific belief revision episode may not erase information because
the other revisions may provide or entail the same information. Whether it does
was proved \conp-hard for sequences of two arbitrary lexicographic revisions or
arbitrarily long lexicographic Horn revisions. A polynomial algorithm is
presented for the case of two lexicographic Horn revision. Heterogeneous
sequences, including revisions other than lexicographic, were proved to belong
in \D2. Their previously proved \conp-hardness is enhanced to \dpi-hardness.

\end{abstract}

\section{Introduction}

Information loss may be the aim of
forgetting~\cite{myer-04,conn-17,gonc-etal-17}, an unintended
consequence~\cite{norb-15,delg-17} or just plainly
unwanted~\cite{delg-wang-15,erde-ferr-07,wang-etal-05,smyt-kean-95,
wang-etal-14,lang-marq-10}. Whichever the case, it is the foremost outcome of
forgetting.

Forgetting a specific revision episode is expected to erase information, but
the same information may be provided or implied by the other revisions. An
algorithm checks whether it does for a heterogeneous sequence of
revisions~\cite{libe-24-a}: lexicographic~\cite{naya-94},
natural~\cite{spoh-88,bout-96-a}, restrained~\cite{boot-meye-06}, very radical,
severe, plain severe, moderate severe~\cite{rott-09} and full
meet~\cite{libe-23}. This \D{2} problem simplifies to \conp-complete for
sequences of lexicographic revisions only~\cite{libe-24-a}.

This complexity analysis misses the hardness of heterogeneous sequences of
revisions, sequences that may contain different kinds of revisions. In this
article, they are proved strictly harder than homogeneous sequences of
lexicographic revisions under the common presumptions of complexity theory.
This result raises the question of why revisions that have the same complexity
of inference~\cite{libe-97-c,libe-23} differ on forgetting. The answer lies in
the complexity of model comparison: sorting two models is easy for
lexicographic revisions and hard for other kinds of revisions.

The previous \conp-hardness is proved for homogeneous sequences of two
lexicographic revisions and arbitrary long sequences of Horn
revisions~\cite{libe-24-a}. Sequences of two lexicographic Horn revisions are
proved polynomial in this article. Their analysis spun two side results:

\begin{itemize}

\item a necessary and sufficient condition to the redundancy of the first
of a sequence of lexicographic revisions from the flat doxastic state;

\item a polynomial-time algorithm for establishing whether a Horn formula is
equivalent to the negation of another Horn formula.

\end{itemize}

An outline of the article follows.
Section~\ref{definitions} sets the framework and provides the necessary
definitions.
The hardness for heterogeneous sequences of revisions is in
Section~\ref{mixed},
the hardness of sorting two models in Section~\ref{comparison}.
Section~\ref{condition} proves a necessary and sufficient condition for
sequences of lexicographic revisions.
Section~\ref{horn} presents a polynomial algorithm for sequences of two Horn
lexicographic revisions.
Section~\ref{conclusions} summarizes and discusses the results in this article.

\section{Definitions}
\label{definitions}

\draft

\obeylines

doxastic state = connected preorder

C = [C(0)..C(m)]

I <=C J iff ...

C(0),C(m) minimal and maximal models
min(A), max(A) minimal and maximal models of A
imin(A), imax(A) their indexes in C

revision change C into Crev(A)
specific revisions: lex(A), sev(A), dsev(A)
defined in terms of Crev(A)

omitted conjunctions
formulae in place of sets of models are their models

flat order 0 = [true]

order Ca = [a, -a]

redundancy of a revision in a sequence from a doxastic state

two sections are only about lexicographic:
lex(A) simplifies to A
0A is 0lex(A), which is [A, A]
0[S1..Sm] is 0[lex(S1)..lex(Sm)]

\enddraft

\draft

old/inductive.tex: inductive definitions of all operators

old/disegni.tex: drawings of all operators

\enddraft

\subsection{Doxastic states}

The current beliefs form the doxastic state. Many belief revision studies
assume it a connected preorder between propositional models, reflecting the
relative strength of beliefs~%
\cite{naya-94,bout-96-a,darw-pear-97,rott-pagn-99,rott-09,boot-meye-11}.
Connected preorders can be represented in many ways, such as the sequence of
their equivalence classes:

\[
C = [C(0), \ldots, C(m)]
\]

An order $C$ sort models $I \leq_C J$ if $I$ belongs to a class preceding $J$
or being the same.

\[
I \leq_C J
\mbox{ ~~~ if and only if ~~~ }
I \in C(i), J \in C(j), i \leq j
\]

The minimal models are the class $C(0)$, the most believed situations. The
maximal models $C(m)${\plural} are the least. The minimal and maximal models of
a formula $A$ are respectively denoted $\min(A)$ and $\max(A)$, the index of
their equivalence classes $\imin(A)$ and $\imax(A)$.

The flat doxastic state $\emptyset$ comprises the class $\emptyset(0)$ of all
models. The doxastic state $C_F$ of a single formula $F$ comprises the class
$C_F(0)$ of the models of $F$ and the class $C_F(1)$ of all others.

\subsection{Notation simplifications}

\begin{description}

\item[Omitted conjunctions.]

In analogy to the omission of multiplication in arithmetic, conjunctions may be
omitted in propositional formulae: $a b (c \vee b)$ stands for $a \wedge b
\wedge (c \vee b)$.

\item[Formulae in place of sets of models.]

If a propositional formula occurs where a set of models is expected, it stands
for its models. For example, the order $[F, \neg F]$ comprises a first class
made of the models of $F$ and a second made of the models of $\neg F$. It is
therefore $C_F$.

\end{description}

\subsection{Belief revision}

A revision is a propositional formula $A$ encoding the new belief. It changes a
doxastic state into a new one. No single revision is universally accepted.
Lexicographic revision, natural revision, radical revision, severe revision,
restrained revision, full meet revision are alternatives.

\begin{description}

\item[Lexicographic revision] introduces new beliefs in all possible
situations, not just the most believed ones~\cite{libe-23}.

\[
C \lex(A) = [
	C(0) \cap A, \ldots, C(m) \cap A,
	C(0) \backslash A, \ldots, C(m) \backslash A
]
\]

\item[Natural revision] trusts the new information $A$ only in the currently
most believed situations $\imin(A)$ where it is true, not in all of
them~\cite{libe-23}.

\begin{eqnarray*}
\lefteqn{C \nat(A)}
\\
&=& [
	\min(A),
	C(0),
	\ldots
	C(\imin(A)-1),
	C(\imin(A)) \backslash A,
\\
&&
	C(\imin(A)+1),
	\ldots,
	C(m)
]
\end{eqnarray*}

\item[Severe revision] does the same, and equates the situations previously
believed more $C(0) \cup \cdots \cup C(\imin(A))$.

\[
C \sev(A) = [
	\min(A),
	C(0) \cup \cdots \cup C(\imin(A)) \backslash A,
	C(\imin(A) + 1), \ldots, C(m)
]
\]

\item[Moderate severe revision] differs on models compatible with the revision
in that it behaves like lexicographic revision.

\begin{eqnarray*}
\lefteqn{C \msev(A)}
\\
&=& [
	C(\imin(A)) \cap A, \ldots, C(\imax(A)) \cap A,
\\
&&
	C(0) \cup \cdots \cup C(\imin(A)) \backslash A,
\\
&&
	C(\imin(A)+1) \backslash A, \ldots, C(m) \backslash A
]
\end{eqnarray*}

\item[Deep severe revision] is a mix of severe and lexicographic revisions: all
models of $A$ precede all models of $\neg A$ like in lexicographic revision and
some classes are merged like in severe revision%
.

\begin{eqnarray*}
\lefteqn{C \dsev(A) =} \\
&&
[
	C(0) \cap A, \ldots, C(m) \cap A,
	(C(0) \cup \cdots \cup C(\imax(A))) \backslash A, \\
&&
	C(\imax(A) + 1), \ldots, C(m)
] 
\end{eqnarray*}

\item[Restrained revision] trusts the new information in the most believed
situations, but otherwise it trusts it minimally.

\begin{eqnarray*}
\lefteqn{C \res(A)}
\\
&=& [
	\min(A),
\\
&&
	C(0) \cap A \backslash \min(A), C(0) \backslash A,
	\ldots
	C(m) \cap A \backslash \min(A), C(m) \backslash A
]
\end{eqnarray*}

\item[Very radical revision] distrusts everything that contradicts the new
information.

\[
C \vrad(A) = [
	C(\imin(A)) \cap A, \ldots, C(\imax(A)) \cap A,
	\true \backslash A
]
\]

\item[Full meet revision] believes the new information only in the most likely
situations where it is true, like natural and severe revisions. At the same
time, it disbelieves everything else equally.

\[
C \full(A) = [
	\min(A),
	\true \backslash \min(A)
]
\]

\end{description}

An arbitrary revision operator is denoted $\rev(A)$.

Revisions in a sequence are applied in order:

\begin{eqnarray*}
C []	&=&
	C	
\\
C [\rev(S_1), \rev(S_2), \ldots, \rev(S_m)]
&=&
	C \rev(S_1) [\rev(S_2), \ldots, \rev(S_m)]
\end{eqnarray*}

\subsection{Redundancy of revisions}

A revision is redundant in a sequence if its removal leads to the same
resulting doxastic state.

\begin{definition}
\label{redundancy}

The revision $S_i$ is redundant in the sequence
{} $[\rev(S_1),\ldots,\rev(S_{i-1}),\rev(S_i),\rev(S_{i+1}),\ldots,\rev(S_m)]$
from the doxastic state $C$ if
{} $C [\rev(S_1),\ldots,\rev(S_{i-1}),\rev(S_i),\rev(S_{i+1}),\ldots,\rev(S_m)]$
coincides with 
{} $C [\rev(S_1),\ldots,\rev(S_{i-1}),\rev(S_{i+1}),\ldots,\rev(S_m)]$.

\end{definition}

The redundant revision do not cause loss of information when forgetting, and
the other way around. For this reason, the redundancy of the first revision in
a sequence is especially relevant: older revisions are more likely to be
forgotten than newer. Redundancy from the flat doxastic state is also
interesting as the flat doxastic state is the plausible case of total initial
ignorance.

Two following sections concern lexicographic revisions only. Notation is
simplified by omitting the revision symbol $\lex()$. For example, $\emptyset A$
is $\emptyset \lex(A)$ and $\emptyset [S_1,\ldots,S_m]$ is $\emptyset
[\lex(S_1),\ldots,\lex(S_m)]$.

\section{Heterogeneous sequences of revisions}
\label{mixed}

\draft

- dpi-hard for natural and severe from non-flat order

- np-hard for deep severe and lexicographic from the flat order

\enddraft

\draft

old/mixed.tex:

summary of relevant results of [forget]

irredundancy expressed as "revision changes, revision revers"

hardness of equivalence, intuition

hardness of redundancy, intuition

\enddraft

The redundancy of heterogenous sequences of revisions is proved \dpi-hard by
the following theorem. This is the problem of checking whether forgetting is
unharmful to information.


\state{hetero-hard}{Theorem}{}{

\begin{theorem}
\label{hetero-hard}

Checking the redundancy of a revision in a heterogeneous sequence of natural
and severe revisions is \dpi-hard.

\end{theorem}

}{

\proof The claim is proved by reduction from the \dpi-hard problem SAT-UNSAT:
establish whether a formula $F$ is satisfiable and another $G$ on a different
alphabet is unsatisfiable~\cite{janc-etal-04}.

The reduction produces an order $C$, a natural revision $\nat(N)$ and a severe
revision $\sev(S)$ such that $\nat(N)$ is redundant in $C \nat(N) \sev(S)$ if
and only if $F$ is satisfiable and $G$ is unsatisfiable.

The keystone of the reduction is that natural revision increases the strength
of belief in certain models $A$ if $F$ is satisfiable and severe revisions
decreases it if $G$ is unsatisfiable.

\long\def\ttytex#1#2{#1}
\ttytex{
}{
                           +-----+                   +-----++-----+
                           |  A  |                   |     ||     |
                           +-----+                   +-----++-----+
                              ^                         |      ^
                              |                         |      |
+--------------+          +---|----------+          +---|------|---+
|              |          |   |          |          |   |      |   |
|              |          |   |          |          |   |      |   |
|              |          |   |          |          |   |      |   |
+--------------+          +---|----------+          |   V      |   |
|+-----++-----+|  nat(N)  |+-----++-----+|  sev(S)  |+-----++-----+|
||  A  ||     ||    =>    ||     ||     ||    =>    ||  A  ||     ||
|+-----++-----+|          |+-----++-----+|          |+-----++-----+|
+--------------+          +--------------+          +--------------+
|              |          |              |          |              |
|              |          |              |          |              |
|              |          |              |          |              |
+--------------+          +--------------+          +--------------+
       C                     C nat(N)               C nat(N) sev(S)
}

The dependency of these changes to the satisfiability of $F$ and $G$ conditions
the revisions.

\begin{itemize}

\item a natural revision only increases the strength of its minimal models;

\item a severe revision only decreases the strength of the models that are
believed as much as its minimal models.

\end{itemize}

The first dependency is achieved if the natural revision comprises the models
of $A$ if and only if $F$ is satisfiable and other models less believed than
them.

The second dependency is achieved if the severe revision contains models more
believed than $A$ if and only if $G$ is satisfiable.

\long\def\ttytex#1#2{#1}
\ttytex{
}{
               +--------------+
               |       +-----+|
               |       |  G  || ---+
               |       +--+--+|    |
               +----------|---+    | S
               |+-----++--+--+|    |
         +---  ||A (F)||     || ---+
         |     |+--+--++-----+|
       N |     +---|----------+
         |     |+--+--+       |
         +---  ||     |       |
               |+-----+       |
               +--------------+
                       C
}

This way, natural revision increases the strength of belief in $F$ and severe
revision decreases it exactly when $F$ is satisfiable and $G$ is unsatisfiable.

\

This mechanism is implemented by the following order and revisions.

\long\def\ttytex#1#2{#1}
\ttytex{
}{
            c     -c
        +--------------+
        |       +----+ |
        |       |a-cG| | --+
        |       +--+-+ | a |
        +----------|---+   | S
        |+-----++--+--+|   |
    +-- ||-abcF||-ab-c|| --+
    |   |+--+--++-----+| -ab
  N |   +---|----------+
    |   |+--+--+       |
    +-- ||-a-bc|       | -a-b
        |+-----+       |
        +--------------+
                C
}

\begin{eqnarray*}
\lefteqn{C \nat(N) \sev(S)}		\\
C &=& [a,  \neg ab,  \neg a \neg b] 	\\ 
N &=& \neg abcF \vee \neg a \neg bc	\\
S &=& a \neg cG \vee \neg ab \neg c
\end{eqnarray*}

The redundancy of $\nat(N)$ is the equivalence of $C \nat(N) \sev(S)$ with $C
\sev(S)$. It depends on the satisfiability of $F$ and $G$.

\begin{description}

\item[$F$ unsatisfiable]\

The first revision
{} $N = \neg abcF \vee \neg a\neg bc$ is equivalent to
{}                    $\neg a\neg bc$.
Its models belong to the third class of $C = [a, \neg a b, \neg a \neg b]$, the
initial order:
{} $\min(N) = N = \neg a \neg bc$
and
{} $\imin(N) = 2$.

\begin{eqnarray*}
\lefteqn{C \nat(N)}
\\
&=& [
	\min(N),
	C(0),
	\ldots
	C(\imin(N)-1),
\\
&&
	C(\imin(N)) \backslash N,
	C(\imin(N)+1),
	\ldots,
	C(m)
]
\\
&=& [
	N,
	C(0),
	\ldots
	C(2-1),
	C(2) \backslash N,
	C(2+1),
	\ldots,
	C(2)
]
\\
&=& [
	N,
	C(0),
	\ldots
	C(1),
	C(2) \backslash N,
	C(3),
	\ldots,
	C(2)
]
\\
&=& [
	N,
	C(0),
	C(1),
	C(2) \backslash N
]
\\
&=& [
	\neg a \neg b c,
	a,
	\neg a b,
	\neg a \neg b \wedge \neg (\neg a \neg b c)
]
\\
&=& [
	\neg a \neg b c,
	a,
	\neg a b,
	\neg a \neg b \neg c
]
\end{eqnarray*}

\long\def\ttytex#1#2{#1}
\ttytex{
}{
                          +-----+
                          |-a-bc|
                          +-----+
+--------------+         +--------------+
|       +----+ |         |       +----+ |
|       |a-cG| | a       |       |a-cG| |
|       +--+-+ |         |       +--+-+ |
+----------|---+         +----------|---+
|       +--+--+|         |       +--+--+|
|       |-ab-c||   =>    |       |-ab-c||
|       +-----+| -ab     |       +-----+|
+--------------+         +--------------+
|+-----+       |         |+-----+       |
||-a-bc|       | -a-b    ||     |       |
|+-----+       |         |+-----+       |
+--------------+         +--------------+
       C                      C nat(N)
}

The minimal models of the following severe revision
{} $S = a \neg c G \vee \neg a b \neg c$
are in the second class $a$ of
{} $C \nat(N) = [\neg a \neg b c, a, \neg a b, \neg a \neg b \neg c]$
if $G$ is satisfiable and in the third $\neg a b$ otherwise. Severe revision
changes no class following its minimal class; it does not change the fourth
class
{} $\neg a \neg b \neg c$
either way.

The minimal models of
{} $S = a \neg c G \vee \neg a b \neg c$
are either in the first or second class of
{} $C = [a,  \neg ab,  \neg a \neg b]$.
Severe revision changes no class following its minimal class; It does not
change the third class
{} $\neg a \neg b$.

The orders differ:
$C \sev(S)$ contains $\neg a \neg b$ while
$C \nat(N) \sev(S)$ contains $\neg a \neg b \neg c$.

The natural revision is irredundant if $F$ is unsatisfiable.

\item[$F$ satisfiable]\

The models of
{} $N = \neg abcF \vee \neg a \neg bc$
are either models of $\neg a b c F$ or models of $\neg a \neg b c$. The former
precede the latter in the order
{} $C = [a, \neg ab, \neg a \neg b]$,
and are therefore minimal:
{} $\min(N) = \neg abcF$
and
{} $\imin(N) = 1$.

\long\def\ttytex#1#2{#1}
\ttytex{
}{
                           +-----+
                           |-abcF|
                           +-----+
+--------------+          +--------------+
|       +----+ |          |       +----+ |
|       |a-cG| | a        |       |a-cG| |
|       +--+-+ |          |       +--+-+ |
+----------|---+          +----------|---+
|+-----++--+--+|          |+-----++--+--+|
||-abcF||-ab-c||    =>    ||     ||-ab-c||
|+--+--++-----+| -ab      |+-----++-----+|
+---|----------+          +--------------+
|+--+--+       |          |+-----+       |
||-a-bc|       | -a-b     ||-a-bc|       |
|+-----+       |          |+-----+       |
+--------------+          +--------------+
       C                      C nat(N)
}

\begin{eqnarray*}
\lefteqn{C \nat(N)}
\\
&=& [
	\min(N),
	C(0),
	\ldots
	C(\imin(N)-1),
\\
&&
	C(\imin(N)) \backslash N,
	C(\imin(N)+1),
	\ldots,
	C(m)
]
\\
&=& [
	\min(N),
	C(0),
	\ldots
	C(1-1),
	C(1) \backslash \min(N),
	C(1+1),
	\ldots,
	C(2)
]
\\
&=& [
	\min(N),
	C(0),
	\ldots
	C(0),
	C(1) \backslash \min(N),
	C(2),
	\ldots,
	C(2)
]
\\
&=& [
	\min(N),
	C(0),
	C(1) \backslash \min(N),
	C(2)
]
\\
&=& [
	\neg abcF,
	a,
	\neg a b \wedge \neg abcF,
	\neg a \neg b
]
\\
&=& [
	\neg abcF,
	a,
	\neg a b \neg (cF),
	\neg a \neg b
]
\end{eqnarray*}

The following severe revision
{} $S = a \neg c G \vee \neg a b \neg c$
depends on the satisfiability of $G$.

\

\begin{description}

\item[$G$ satisfiable]\

The severe revision
{} $S = a \neg c G \vee \neg a b \neg c$
comprises the models of $a \neg c G$ and the models of $\neg a b \neg c$.
The former precede the latter in the order
{} $D = C \nat(N) = [\neg abcF, a, \neg a b \neg (cF), \neg a \neg b]$
and are therefore minimal:
{} $\min(S) = a \neg c G$
amd
{} $\imin(S) = 1$.

\long\def\ttytex#1#2{#1}
\ttytex{
}{
                           +-----+                          +----+
                           |-abcF|                          |a-cG|
                           +-----+                          +----+
+--------------+          +--------------+          +--------------+
|       +----+ |          |       +----+ |          |       +----+ +-----+
|       |a-cG| | a        |       |a-cG| |          |       |    | |-abcF|
|       +--+-+ |          |       +--+-+ |          |       +----+ +-----+
+----------|---+          +----------|---+          +--------------+
|+-----++--+--+|          |+-----++--+--+|          |+-----++-----+|
||-abcF||-ab-c||    =>    ||     ||-ab-c||    =>    ||     ||-ab-c||
|+--+--++-----+| -ab      |+-----++-----+|          |+-----++-----+|
+---|----------+          +--------------+          +--------------+
|+--+--+       |          |+-----+       |          |+-----+       |
||-a-bc|       | -a-b     ||-a-bc|       |          ||-a-bc|       |
|+-----+       |          |+-----+       |          |+-----+       |
+--------------+          +--------------+          +--------------+
       C                    D = C nat(N)         D sev(S) = C nat(N) sev(S)
}

\begin{eqnarray*}
\lefteqn{C \nat(N) \sev(S)}
\\
&=& D \sev(S)
\\
&=& [
	\min(S),
	D(0) \cup \cdots \cup D(\imin(S)) \backslash S,
	D(\imin(S) + 1),
	\ldots,
	D(m)
]
\\
&=& [
	\min(S),
	D(0) \cup \cdots \cup D(1) \backslash S,
	D(1 + 1),
	\ldots,
	D(3)
]
\\
&=& [
	\min(S),
	D(0) \cup D(1) \backslash S,
	D(2),
	D(3)
]
\\
&=& [
	\min(S),
	D(0) \cup D(1) \backslash S,
	\neg a b \neg (cF),
	D(3)
]
\end{eqnarray*}

\draft

the remaining steps are unnecessary:

\begin{eqnarray*}
\lefteqn{C \nat(N) \sev(S)}
\\
&=& D \sev(S)
\\
&=& [
	\min(S),
	D(0) \cup \cdots \cup D(\imin(S)) \backslash S,
	D(\imin(S) + 1),
	\ldots,
	D(m)
]
\\
&=& [
	\min(S),
	D(0) \cup \cdots \cup D(1) \backslash S,
	D(1 + 1),
	\ldots,
	D(3)
]
\\
&=& [
	\min(S),
	D(0) \cup D(1) \backslash S,
	D(2),
	D(3)
]
\\
&=& [
	\min(S),
	(D(0) \backslash S) \cup (D(1) \backslash S),
	D(2),
	D(3)
]
\\
&=& [
	\min(S),
	D(0) \cup D(1) \backslash \min(S),
	D(2),
	D(3)
]
\\
&=& [
	\min(S),
	D(0) \backslash \min(S) \cup D(1),
	D(2),
	D(3)
]
\\
&=& [
	a \neg c G,
	a \neg c \backslash a \neg c G \cup D(1),
	D(2),
	D(3)
]
\\
&=& [
	a \neg c G,
	a \neg c \neg G \cup D(1),
	D(2),
	D(3)
]
\end{eqnarray*}

\enddraft

The models of $\neg a b$ are split in class $\neg a b \neg (cF)$ and another
class. This is the effect of the previous natural revision, which is therefore
irredundant.

Removing $\nat(N)$ from $C \nat(N) \sev(S)$ leaves only $C \sev(S)$. The
minimal models of
{} $S = a \neg c G \vee \neg a b \neg c$
in the order
{} $C = [a, \neg a b, \neg a \neg b]$
are $\min(S) = a \neg c G$, their index $\imin(S) = 0$.

\long\def\ttytex#1#2{#1}
\ttytex{
}{
                                  +----+ 
                                  |a-cG| 
                                  +----+ 
+--------------+          +--------------+
|       +----+ |          |       +----+ |
|       |a-cG| | a        |       |    | |
|       +--+-+ |          |       +----+ |
+----------|---+          +--------------+
|+-----++--+--+|          |+-----++-----+|
||-abcF||-ab-c||    =>    ||-abcF||-ab-c||
|+--+--++-----+| -ab      |+-----++-----+|
+---|----------+          +--------------+
|+--+--+       |          |+-----+       |
||-a-bc|       | -a-b     ||-a-bc|       |
|+-----+       |          |+-----+       |
+--------------+          +--------------+
       C                      C sev(S)
}

\begin{eqnarray*}
\lefteqn{C \sev(S)}
\\
&=& [
	\min(S),
	C(0) \cup \cdots \cup C(\imin(S)) \backslash S,
	C(\imin(S) + 1),
	\ldots,
	C(m)
]
\\
&=& [
	\min(S),
	C(0) \cup \cdots \cup C(0) \backslash S,
	C(0 + 1),
	\ldots,
	C(2)
]
\\
&=& [
	\min(S),
	C(0) \backslash S,
	C(1),
	C(2)
]
\\
&=& [
	\min(S),
	C(0) \backslash S,
	\neg a b,
	C(2)
]
\end{eqnarray*}

\draft

the remaining steps are unnecessary:

\begin{eqnarray*}
\lefteqn{C \sev(S)}
\\
&=& [
	\min(S),
	C(0) \cup \cdots \cup C(\imin(S)) \backslash S,
	C(\imin(S) + 1),
	\ldots,
	C(m)
]
\\
&=& [
	a \neg c G,
	C(0) \cup \cdots \cup C(0) \backslash (a \neg c G \vee \neg a b \neg c),
	C(0 + 1),
	\ldots,
	C(2)
]
\\
&=& [
	a \neg c G,
	C(0) \backslash (a \neg c G \vee \neg a b \neg c),
	C(1), C(2)
]
\\
&=& [
	a \neg c G,
	a \backslash (a \neg c G \vee \neg a b \neg c),
	\neg a b,
	\neg a \neg b
]
\\
&=& [
	a \neg c G,
	a \backslash \neg (c G),
	\neg a b,
	\neg a \neg b
]
\end{eqnarray*}

\enddraft

The models of $\neg a b$ are all in the third class of $C \sev(S)$, contrary to
$C \nat(N) \sev(S)$, which splits out $\neg a b \neg (cF)$. The latter is not
empty since it contains all models of $\neg a b \neg c$.

The natural revision is irredundant.

\

\item[$G$ unsatisfiable]\

The severe revision
{} $S = a \neg c G \vee \neg a b \neg c$
is equivalent to
{}                     $\neg a b \neg c$.
Its models are all in the third class
{} $D(2) = \neg a b \neg (cF)$
of
{} $D = C \nat(N) = [\neg abcF, a, \neg a b \neg (cF), \neg a \neg b]$.
They are all minimal:
{} $\min(S) = \neg a b \neg c$
and
{} $\imin(S) = 2$.

\long\def\ttytex#1#2{#1}
\ttytex{
}{
                           +-----+                          +-----+
                           |-abcF|                          |-ab-c|
                           +-----+                          +-----+
+--------------+          +--------------+          +--------------+
|              |          |              |          |              |
|              | a        |              |          |              |
|              |          |              |          |              |
+--------------+          +--------------+          |              |
|+-----++-----+|          |+-----++-----+|          |+-----++-----++-----+
||-abcF||-ab-c||    =>    ||     ||-ab-c||    =>    ||     ||     ||-abcF|
|+--+--++-----+| -ab      |+-----++-----+|          |+-----++-----++-----+
+---|----------+          +--------------+          +--------------+
|+--+--+       |          |+-----+       |          |+-----+       |
||-a-bc|       | -a-b     ||-a-bc|       |          ||-a-bc|       |
|+-----+       |          |+-----+       |          |+-----+       |
+--------------+          +--------------+          +--------------+
       C                    D = C nat(N)         D sev(S) = C nat(N) sev(S)
}

\begin{eqnarray*}
\lefteqn{D \sev(S)}
\\
&=& [
	\min(S),
	D(0) \cup \cdots \cup D(\imin(S)) \backslash S,
	D(\imin(S) + 1),
	\ldots,
	D(m)
]
\\
&=& [
	\min(S),
	D(0) \cup \cdots \cup D(2) \backslash S,
	D(2 + 1),
	\ldots,
	D(3)
]
\\
&=& [
	\min(S),
	D(0) \cup \cdots \cup D(2) \backslash S,
	D(3),
	D(3)
]
\\
&=& [
	\min(S),
	D(0) \cup D(1) \cup D(2) \backslash S,
	D(3)
]
\\
&=& [
	\min(S),
	D(1) \cup D(0) \cup D(2) \backslash S,
	D(3)
]
\\
&=& [
	\neg a b \neg c,
	(a \vee (\neg a b c F) \vee (\neg a b \neg(c F)))
		\wedge 
	\neg (\neg a b \neg c),
	\neg a \neg b
]
\\
&=& [
	\neg a b \neg c,
	(a \vee \neg a b)
		\wedge 
	\neg (\neg a b \neg c),
	\neg a \neg b
]
\end{eqnarray*}

The effect of the natural revision is erased: $\neg a b c F$ is no longer split
out from $\neg a b$. Removing it produces the same order:
{} $C \nat(N) \sev(S) = C \sev(S)$.

The minimal models of
{} $S = a \neg c G \vee \neg a b \neg c =
{}                      \neg a b \neg c$
in the order
{} $C = [a, \neg a b, \neg a \neg b]$
are
$\min(S) = \neg a b \neg c$,
of index
$\imin(S) = 1$.

\long\def\ttytex#1#2{#1}
\ttytex{
}{
                                  +-----+
                                  |-ab-c|
                                  +-----+
+--------------+          +--------------+
|              |          |              |
|              | a        |              |
|              |          |              |
+--------------+          |              |
|+-----++-----+|          |+-----++-----+|
||-abcF||-ab-c||    =>    ||-abcF||     ||
|+--+--++-----+| -ab      |+-----++-----+|
+---|----------+          +--------------+
|+--+--+       |          |+-----+       |
||-a-bc|       | -a-b     ||-a-bc|       |
|+-----+       |          |+-----+       |
+--------------+          +--------------+
       C                      C sev(S)
}

\begin{eqnarray*}
\lefteqn{C \sev(S)}
\\
&=& [
	\min(S),
	C(0) \cup \cdots \cup C(\imin(S)) \backslash S,
	C(\imin(S) + 1), \ldots, C(m)
]
\\
&=& [
	\min(S),
	C(0) \cup \cdots \cup C(1) \backslash S,
	C(1 + 1), \ldots, C(2)
]
\\
&=& [
	\min(S),
	C(0) \cup C(1) \backslash S,
	C(2)
]
\\
&=& [
	\neg a b \neg c,
	(a \vee \neg a b) \wedge \neg (\neg a b \neg c),
	\neg a \neg b
]
\end{eqnarray*}

This order $C \sev(S)$ is identical to $D \sev(S) = C \nat(N) \sev(S)$.

This is the only case where the natural revision is redundant, and is the case
when $F$ is unsatisfiable and $G$ is unsatisfiable.

The redundancy of natural revision when followed by a severe revision is 
\dpi-hard~\cite{janc-etal-04}.

\end{description}

\end{description}
~\qed

}

The proof involves only the redundancy of a single natural revision followed by
a single severe revision. Redundancy is \dpi-hard even in this very simple
case.

The initial order is not flat, the relevant case of total ignorance. Redundancy
from the flat doxastic state is at least \np-hard.

\state{hetero-flat-redundancy}{Theorem}{}{

\begin{theorem}
\statelabel

Checking the redundancy of a revision in a heterogeneous sequence of deep
severe and lexicographic revisions from the flat doxastic state is \np-hard.

\end{theorem}

}{

\proof The proof reduces the satisfiability of a formula $F$ to the redundancy
of $\lex(a \vee b)$ in
{} $[\lex(a \vee b), \lex(a), \sev(D)]$
from the flat doxastic state $\emptyset$, where
{} $D = a \vee (\neg a \neg b c F)$.

The result of the first revision is
{} $\emptyset [\lex(a \vee b)] = [a \vee b, \neg a \neg b]$.

Its first class $a \vee b$ contains the second revision $a$, and its cut is
therefore the only change:
{} $\emptyset [\lex(a \vee b) \lex(a)] =
{}	[a \vee b, \neg a \neg b] \lex(a) =
{}	[a, (a \vee b) \wedge \neg a, \neg a \neg b] =
{}	[a, \neg a b, \neg a \neg b]$.
This order is named $C$.

The models of the third revision $D = a \vee (\neg a \neg b c F)$ comprise
$C(0) = a$ and a part of $C(2) = \neg a \neg b$, the latter only if $F$ is
satisfiable. As a result, $\imax(D)$ is $2$ if $F$ is satisfiable and $0$
otherwise.

\

The order changes as follows if $F$ is satisfiable.

\begin{eqnarray*}
\lefteqn{C \dsev(D) =}
\\
&=&
	[C(0) \cap D, C(2) \cap D, (C(0) \cup C(1) \cup C(2)) \backslash D]
\\
&=& [
	a (a \vee (\neg a \neg b c F)),
	\neg a \neg b (a \vee (\neg a \neg b c F)),
	\true \neg D
]
\\
&=& [
	a,
	\neg a \neg b (\neg a \neg b c F),
	\neg D
]
\\
&=& [
	a,
	\neg a \neg b c F,
	\neg D
]
\end{eqnarray*}

The first revision is redundant as shown by the result of the others only. It
changes the flat order to
{} $0 \lex(a) = [a, \neg a]$.
The maximal class of models of $D$ in this ordering is $\neg a$ since
{} $\neg a D$
is
{} $\neg a (a \vee (\neg a \neg b c F))$,
which is equivalent to
{} $\neg a (\neg a \neg b c F)$
and to
{} $\neg a \neg b c F$,
which is consistent because $F$ is assumed so. This equivalence
{} $\neg a D \equiv \neg a \neg b c F$
is employed in the following chain of equivalences as well.

\begin{eqnarray*}
\lefteqn{\emptyset \lex(a) \dsev(D) =}
\\
&=&
	[a, \neg a] \dsev(D)
\\
&=&
	[a D, \neg a D, (a \vee \neg a) \neg D]
\\
&=& [
	a (a \vee (\neg a \neg b c F)),
	\neg a \neg b c F,
	\neg D
]
\\
&=& [
	a,
	\neg a \neg b c F,
	\neg D
]
\end{eqnarray*}

The equality with 
{} $0 [\lex(a v b), \lex(a), \dsev(D)]$
proves redundancy.

\

If $F$ is unsatisfiable, it is equivalent to $\false$.\
Therefore,
{} $D = a \vee (\neg a \neg b c F)$
is equivalent to
{} $a \vee (\neg a \neg b c \false)$,
which is equivalent to $a$.

The third revision $\dsev(D)$ does not change
{} $C = \emptyset [\lex(a \vee b), \lex(a)] =
{}	[a, \neg a b, \neg a \neg b]$
since $\imax(D) = \imax(a) = 0$.

\begin{eqnarray*}
\lefteqn{C \dsev(D) =}
\\
&=&
	[a D, a \neg D, \neg a b, \neg a \neg b]
\\
&=&
	[a a, a \neg a, \neg a b, \neg a \neg b]
\\
&=&
	[a, \emptyset, \neg a b, \neg a \neg b]
\\
&=&
	[a, \neg a b, \neg a \neg b]
\end{eqnarray*}

The deep severe revision $\dsev(a)$ does not change the order resulting from
the second revision only also.

\begin{eqnarray*}
\lefteqn{\emptyset \lex(a) \dsev(D) =}
\\
&=& [
	a,
	\neg a
] \dsev(a)
\\
&=& [
	a a,
	\neg a \neg a
]
\\
&=& [
	a,
	\neg a
]
\end{eqnarray*}

The difference between orderings disproves the equivalence between
{} $[\lex(a \vee b), \lex(a), \dsev(D)]$
and
{} $[\lex(a), \dsev(D)]$
and the redundancy of $\lex(a)$ in
{} $[\lex(a), \dsev(D)]$.~\qed

}

Redundancy is no-loss forgetting: forgetting a redundant revision does not
alter the doxastic state.

\begin{corollary}
\label{hetero-forget}

Checking whether forgetting a revision in a sequence causes no information loss
is \dpi-hard.

\end{corollary}

\begin{corollary}
\label{hetero-flat-forget}

Checking whether forgetting a revision in a sequence from the flat doxastic
state causes no information loss is \np-hard.

\end{corollary}

\section{Complexity of comparison}
\label{comparison}

Different revisions have different complexity of redundancy, the resilience to
forgetting. Lexicographic revision is \conp-complete. Deep severe and natural
revisions are \np-hard, in \D2, natural and severe revisions are \dpi-hard, in
\D2: they are not in \conp, according to complexity theory prevailing credence.

They have the same complexity of inference, checking whether a formula is
believed true: \D2-complete~\cite{libe-97-c,libe-23}.

Why the difference?

Inference appears the basic problem of iterated revision: representing beliefs.
It tells whether something is believed. Redundancy is a constructed matter: a
formula removed from a sequence does not change the resulting beliefs. Why the
same cost of the basic problem and the differing cost of the constructed?

Complexity itself suggest an answer.

Inference from a doxastic state is truth in all believed models. Validity in
propositional logic is truth in all models. It is \conp-hard. Its complexity
covers the different complexity of redundancy.

Redundancy is not only about the most believed models. It is about the strength
of belief of all models. It is about the whole doxastic state, and doxastic
states are connected preorders. The basic problem of iterated belief revision
is not inference. It is calculating the final order, how it sorts models.

The cost of $I \leq_{C [\rev(S_1),\ldots,\rev(S_m)]} J$ differs between
revision operators: it is polynomial for lexicographic and very radical
revision, \np-hard for deep severe and full meet revisions, \conp-hard for
moderate severe revision, \dpi-hard for natural, severe and restrained
revisions. Polynomial for some operators, hard for others. The differences
reflect on the differences of redundancy.

\subsection{Natural, severe and restrained revisions}

A single reduction proves \dpi-hard all revisions having three properties
possessed by natural, severe and restrained revision. Most operators have two
of them: the flat order revised by $A$ is $C_A = [A,\neg A]$; revising a
two-class order by its second class swaps the two. Many operators do not have
the other property: a revision consistent with the first class of the order
does not invalidate any strict order. Lexicographic, moderate severe and deep
severe revisions invalidate $I < J$ when $J$ satisfies the revision and $I$
does not. Very radical and full meet revisions invalidate $I < J$ when $I$ and
$J$ falsify the revision.

\state{natural-severe-restrained-translation}{Lemma}{}{

\begin{lemma}
\statelabel

A polynomial translation produces a sequence of revisions $S$ and two models
$I$ and $J$ such that $I \leq_{\emptyset S} J$ holds if and only if a formula
$F$ is satisfiable and another $G$ on a disjoint alphabet is unsatisfiable, if
the revision has the following properties.

\begin{enumerate}

\item
{} $\emptyset \rev(A) = C_A$.

\item
{} $C_{\neg A} \rev(A) = C_A$.

\item If $A \cap C(0)$ is not empty, then $I<_CJ$ implies
$I <_{C \rev(A)} J$.

\end{enumerate}

\end{lemma}

}{

\proof The claim is proved by linking $I \leq_{\emptyset S} J$ to both the
satisfiability of a formula $F$ and the unsatisfiability of $G$. Two fresh
variables $a$ and $b$ are required.

\begin{eqnarray*}
I	&=&	\{b\}				\\
J	&=&	\{a\}				\\
S	&=&	[a, \neg a \vee abF, a \vee \neg a \neg b G]
\end{eqnarray*}

The keystone of the proof is the distribution of models of the second and third
revisions in the doxastic state the first produces: $abF$ and $J$ are in the
first class, $I$ and $-a-bG$ in the second.

\long\def\ttytex#1#2{#1}
\ttytex{
}{
+-------------+
|+---+        |
||abF| J      | a
|+---+        |
+--+------+---+
|      +--+--+|
|    I |-a-bG|| -a
|      +-----+|
+-------------+
}

If $F$ is unsatisfiable, $a b F$ has no models. The revision $\neg a \vee a b
F$ is equivalent to $\neg a$. It swaps the classes by the second revision
property, making $I$ strictly less than $J$.

\long\def\ttytex#1#2{#1}
\ttytex{
}{
+-------------+          +-------------+
|             |          |      +-----+|
|      J      | a        |    I |-a-bG|| -a
|             |          |      +-----+|
+-------------+    =>    +-------------+
|      +-----+|          |             |
|    I |-a-bG|| -a       |      J      | a
|      +-----+|          |             |
+-------------+          +-------------+
}

If $G$ is satisfiable, the last revision $a \vee \neg a \neg b G$ is consistent
with the first class $\neg a$. As a result, it maintains $I$ strictly less than
$J$ by the third revision property.

All other satisfiability conditions either swap $I$ and $J$ again or never swap
them at all. The claim is now formally proved.

\

The first revision $a$ changes the flat doxastic state $\emptyset$ into $C_a =
[a,\neg a]$ by the first revision property.

This order $C_a = \emptyset [a]$ is further revised by $S_2 = \neg a \vee abF$
and then by $S_3 = a \vee \neg a \neg b G$. The first depends on the
satisfiability of $F$.

\begin{description}

\item[$F$ satisfiable]\

The second revision $S_2 = \neg a \vee abF$ is consistent with the first class
of $C_a = [a, \neg a]$. It maintains every model of $a$ strictly less than
every model of $\neg a$ by the third revision property. Three consequences are:

\begin{itemize}

\item $J \in a$ is strictly less than $I \in \neg a$;

\item class zero does not contain all models,
as that would make $J$ equal to $I$ instead of strictly less;

\item class zero contains no model of $\neg a$,
as that would make $J \in a$ no longer strictly less than that model.

\end{itemize}

\long\def\ttytex#1#2{#1}
\ttytex{
}{
                          +---+
                          |abF|
                          +---+
+-------------+          +-------------+
|+---+        |          |+---+        |
||abF| J      | a        ||///| J      | a - abF
|+-+-+        |          |+---+        |
+--+------+---+    =>    +-------------+
|      +--+--+|          |      +-----+|
|    I |-a-bG|| -a       |    I |-a-bG|| -a
|      +-----+|          |      +-----+|
+-------------+          +-------------+
}

Since the first class does not include models of $\neg a$ and is not empty by
definition, it contains models of $a$. It is therefore consistent with the last
revision $S_3 = a \vee \neg a \neg b G$ regardless of the satisfiability of
$G$.

By the third revision property, $J$ remains strictly less than $I$ regardless
of the satisfiability of $G$.

\item[$F$ is unsatisfiable]\

The revision $S_2 = \neg a \vee abF$ is equivalent to $\neg a$, the second
class of the order $C_a = [a, \neg a]$. It swaps the classes by the second
revision property: $C_a \rev(S_2) = [\neg a, a] = C_{\neg a}$.

\long\def\ttytex#1#2{#1}
\ttytex{
}{
+-------------+          +-------------+
|             |          |      +-----+|
|      J      | a        |    I |-a-bG|| -a
|             |          |      +-----+|
+-------------+    =>    +-------------+
|      +-----+|          |             |
|    I |-a-bG|| -a       |      J      | a
|      +-----+|          |             |
+-------------+          +-------------+   
}

The sequel depends on the satisfiability of $G$.

\begin{description}

\item[$G$ is unsatisfiable]\

The conjunction $\neg a \neg b G$ has no models. The third revision
{} $S_3 = a \vee \neg a \neg b G$
is equivalent to $a$, exactly the second class of the order
$C_a \rev(S_2) = C_{\neg a} = [\neg a,a]$.
Therefore, it swaps the classes again:
$C_a \rev(S_2) \rev(S_3) = C_{\neg a} \rev(S_3) = C_{\neg a} \rev(a) =  C_a$.

\long\def\ttytex#1#2{#1}
\ttytex{
}{
+-------------+          +-------------+
|             |          |             |
|    I        | -a       |      J      | a
|             |          |             |
+-------------+    =>    +-------------+
|             |          |             |
|      J      | a        |    I        | -a
|             |          |             |
+-------------+          +-------------+
}

Consequently, $I$ and $J$ are swapped again: $J \in a$ is strictly less than $I
\in \neg a$.

\item[$G$ is satisfiable]\

The last revision $S_3 = a \vee \neg a \neg b G$ is consistent with the first
class $\neg a$ of $C_a \rev(S_2) = C_{\neg a}$. It maintains $I \in \neg a$
strictly less than $J \in a$.

\long\def\ttytex#1#2{#1}
\ttytex{
}{
				+-----+
				|-a-bG|
				+-----+
+-------------+          +-------------+
|      +-----+|          |      +-----+|
|    I |-a-bG|| -a       |    I |XXXXX|| -a -(-a-bG)
|      +-----+|          |      +-----+|
+-------------+    =>    +-------------+
|             |          |             |
|      J      | a        |      J      | a
|             |          |             |
+-------------+          +-------------+
}

These are the only conditions resulting in $I < J$: $F$ is unsatisfiable and
$G$ is satisfiable. All others result in $J < I$. This proves that $I \leq_S J$
holds if and only if $F$ is unsatisfiable and $G$ is satisfiable.

\end{description}

\end{description}
\qed

}

Natural, severe and restrained revisions possess the required three properties.

\state{natural-properties}{Lemma}{}{

\begin{lemma}
\statelabel

Natural revision satisfies the following properties.

\begin{enumerate}

\item
{} $\emptyset \nat(A) = C_A$.

\item
{} $[C(0),C(1)] \nat(C(1)) = [C(1),C(0)]$.

\item If $A \cap C_0$ is not empty, then $I<_CJ$ implies
$I <_{C \rev(A)} J$.

\end{enumerate}

\end{lemma}

}{

\proof
\begin{enumerate}


\item
The minimal index of $A$ in $\emptyset$ is 0, its minimal models $A$.

\begin{eqnarray*}
\emptyset \nat(A)
\\
&=&
[
	\min(A),
	\emptyset(0),
	\ldots
	\emptyset(\imin(A)-1),
\\
&&
	\emptyset(\imin(A)) \backslash A,
	\emptyset(\imin(A)+1),
	\ldots,
	\emptyset(m)
]
\\
&=&
[
	A,
	\emptyset(0),
	\ldots
	\emptyset(0-1),
	\emptyset(0) \backslash A,
	\emptyset(0+1),
	\ldots,
	\emptyset(0)
]
\\
&=&
[
	A,
	\emptyset(0) \backslash A
]
\\
&=&
[
	A,
	\neg A
]
\\
&=&
	C_A
\end{eqnarray*}


\item

The minimal models of $C(1)$ in $[C(0),C(1)]$ are $C(1)$, their index $1$.

\begin{eqnarray*}
\lefteqn{C \nat(C(i))}
\\
&=& [
	\min(C(1)),
	C(0),
	\ldots
	C(\imin(C(1))-1),
\\
&&
	C(\imin(C(1))) \backslash C(1),
	C(\imin(C(1))+1),
	\ldots,
	C(m)
]
\\
&=& [
	C(1),
	C(0),
	\ldots
	C(1-1),
	C(1) \backslash C(1),
	C(1+1),
	\ldots,
	C(1)
]
\\
&=& [
	C(1),
	C(0),
	\ldots
	C(0),
	C(2),
	\ldots,
	C(1)
]
\\
&=& [
	C(1),
	C(0)
]
\end{eqnarray*}


\item

Since $A \cap C_0$ is not empty, the minimal models of $A$ are $C(0) \cap A$,
its minimal index is $0$.

\begin{eqnarray*}
\lefteqn{C \nat(A)}
\\
&=& [
	\min(A),
	C(0),
	\ldots
	C(\imin(A)-1),
\\
&&
	C(\imin(A)) \backslash A,
	C(\imin(A)+1),
	\ldots,
	C(m)
]
\\
&=& [
	C(0) \cap A
	C(0),
	\ldots
	C(0-1),
	C(0) \backslash A,
	C(0+1),
	\ldots,
	C(m)
]
\\
&=& [
	C(0) \cap A,
	C(0) \backslash A,
	C(1),
	\ldots,
	C(m)
]
\end{eqnarray*}

If $I$ is strictly less than $J$ in $C$, they belong respectively to $C(i)$ and
$C(j)$ with $i<j$.

If $i=0$ then $j>0$. Therefore, $I$ is either in $C(0) \cap A$ or in $C(0)
\backslash A$ and $J$ is in $C(1) \cup \cdots \cup C(m)$>

If $i > 0$ then $j > 0$. Both $I$ and $J$ are in classes among
$C(1),\ldots,C(m)$. These class do not change order.

\end{enumerate}
\qed

}

\state{severe-properties}{Lemma}{}{

\begin{lemma}
\statelabel

Severe revision satisfies the following properties.

\begin{enumerate}

\item
{} $\emptyset \sev(A) = C_A$.

\item
{} $[C(0),C(1)] \sev(C(1)) = [C(1),C(0)]$.

\item If $A \cap C_0$ is not empty, then $I<_CJ$ implies
$I <_{C \sev(A)} J$.

\end{enumerate}

\end{lemma}

}{

\proof
\begin{enumerate}


\item

The minimal models of $A$ in $\emptyset$ are $A$, their index $0$.

\begin{eqnarray*}
\lefteqn{C \sev(A)}
\\
&=& [
	\min(A),
	C(0) \cup \cdots \cup C(\imin(A)) \backslash A,
	C(\imin(A) + 1), \ldots, C(m)
]
\\
&=& [
	A,
	C(0) \cup \cdots \cup C(0) \backslash A,
	C(0 + 1), \ldots, C(0)
]
\\
&=& [
	A,
	C(0) \backslash A,
]
\\
&=& [
	A,
	\true \backslash A,
]
\\
&=& [
	A,
	\neg A,
]
\\
&=&
	C_A
\end{eqnarray*}


\item

The minimal models of $C(1)$ in $[C(0),C(1)]$ are $C(1)$, their index $1$.

\begin{eqnarray*}
\lefteqn{C \sev(C(1))}
\\
&=& [
	\min(C(1)),
	C(0) \cup \cdots \cup C(\imin(A)) \backslash C(1),
	C(\imin(A) + 1), \ldots, C(m)
]
\\
&=& [
	C(1),
	C(0) \cup \cdots \cup C(1) \backslash C(1),
	C(1 + 1), \ldots, C(1)
]
\\
&=& [
	C(1),
	C(0) \cup C(1) \backslash C(1),
	C(2), \ldots, C(1)
]
\\
&=& [
	C(1),
	C(0)
]
\end{eqnarray*}


\item

Since $A \cap C(0)$ is not empty, the minimal models of $A$ in $C$ are $C(0)
\cap A$, their index 0.

\begin{eqnarray*}
\lefteqn{C \sev(A)}
\\
&=& [
	\min(A),
	C(0) \cup \cdots \cup C(\imin(A)) \backslash A,
	C(\imin(A) + 1), \ldots, C(m)
]
\\
&=& [
	C(0) \cap A,
	C(0) \cup \cdots \cup C(0) \backslash A,
	C(0 + 1), \ldots, C(m)
]
\\
&=& [
	C(0) \cap A,
	C(0) \backslash A,
	C(1), \ldots, C(m)
]
\end{eqnarray*}

If $I$ is strictly less than $J$ in $C$, they belong respectively to $C(i)$ and
$C(j)$ with $i<j$.

If $i=0$ then $j>0$. Therefore, $I$ is either in $C(0) \cap A$ or in $C(0)
\backslash A$ and $J$ is in $C(1) \cup \cdots \cup C(m)$>

If $i > 0$ then $j > 0$. Both $I$ and $J$ are in classes among
$C(1),\ldots,C(m)$. These class do not change order.

\end{enumerate}
\qed

}

\state{restrained-properties}{Lemma}{}{

\begin{lemma}
\statelabel

Restrained revision satisfies the following properties.

\begin{enumerate}

\item
{} $\emptyset \res(A) = C_A$.

\item
{} $[C(0),C(1)] \res(C(1)) = [C(1),C(0)]$.

\item If $A \cap C_0$ is not empty, then $I<_CJ$ implies
$I <_{C \res(A)} J$.

\end{enumerate}

\end{lemma}

}{

\proof
\begin{enumerate}


\item

The minimal models of $A$ in $\emptyset$ are $A$.

\begin{eqnarray*}
\lefteqn{C \res(A)}
\\
&=& [
	\min(A),
	C(0) \cap A \backslash \min(A), C(0) \backslash A,
	\ldots
	C(m) \cap A \backslash \min(A), C(m) \backslash A
]
\\
&=& [
	A,
	C(0) \cap A \backslash A, C(0) \backslash A,
	\ldots
	C(0) \cap A \backslash A, C(0) \backslash A
]
\\
&=& [
	A,
	C(0) \cap A \backslash A, C(0) \backslash A,
]
\\
&=& [
	A,
	\emptyset, \true \backslash A,
]
\\
&=& [
	A,
	\neg A
]
\end{eqnarray*}


\item

The minimal models of $C(1)$ in $[C(0),C(1)]$ are $C(1)$.

\begin{eqnarray*}
\lefteqn{C \res(C(1))}
\\
&=& [
	\min(C(1)),
	C(0) \cap C(1) \backslash \min(C(1)), C(0) \backslash C(1),
\\
&&
	\ldots
	C(m) \cap C(1) \backslash \min(C(1)), C(m) \backslash C(1)
]
\\
&=& [
	C(1),
	C(0) \cap C(1) \backslash C(1), C(0) \backslash C(1),
	\ldots
	C(1) \cap C(1) \backslash C(1), C(1) \backslash C(1)
]
\\
&=& [
	C(1),
	C(0) \cap C(1) \backslash C(1), C(0) \backslash C(1),
	C(1) \cap C(1) \backslash C(1), C(1) \backslash C(1)
]
\\
&=& [
	C(1),
	\emptyset, C(0) \backslash C(1),
	\emptyset, \emptyset
]
\\
&=& [
	C(1),
	C(0)
]
\end{eqnarray*}


\item

Since $C(0) \cap A$ is not empty, the minimal models of $A$ are $C(0) \cap A$.

\begin{eqnarray*}
\lefteqn{C \res(A)}
\\
&=& [
	\min(A),
	C(0) \cap A \backslash \min(A), C(0) \backslash A,
	\ldots
	C(m) \cap A \backslash \min(A), C(m) \backslash A
]
\\
&=& [
	C(0) \cap A,
	C(0) \cap A \backslash C(0) \cap A, C(0) \backslash A,
\\
&&
	C(1) \cap A \backslash C(0) \cap A, C(1) \backslash A,
	\ldots
	C(m) \cap A \backslash C(0) \cap A, C(m) \backslash A
]
\\
&=& [
	C(0) \cap A,
	\emptyset, C(0) \backslash A,
	C(1) \cap A, C(1) \backslash A,
	\ldots
	C(m) \cap A, C(m) \backslash A
]
\\
&=& [
	C(0) \cap A, C(0) \backslash A,
	C(1) \cap A, C(1) \backslash A,
	\ldots
	C(m) \cap A, C(m) \backslash A
]
\end{eqnarray*}

If $I$ is strictly less than $J$ in $C$, then $I$ is $C(i)$ and $J$ in $C(j)$
with $i < j$. As a result, $I$ is in either $C(i) \cap A$ or $C(i) \backslash
A$ and $J$ is in either $C(j) \cap A$ or $C(j) \backslash A$. Both $C(i) \cap
A$ and $C(i) \backslash A$ precede both $C(j) \cap A$ or $C(j) \backslash A$ in
$C \res(A)$. Therefore, $I <_{C \res(A)} J$.

\end{enumerate}
\qed

}

Since natural, severe and restrained revision possess the properties required
by the reduction, they are \dpi-hard.

\begin{corollary}
\label{natural-severe-restrained-hard}

Establishing $I \leq_{\emptyset S} J$ is \dpi-hard if $S$ is a sequence of
natural revisions or a sequence of severe revisions or a sequence of restrained
revisions.

\end{corollary}

\subsection{Full meet revision}

The reduction of Lemma~\ref{natural-severe-restrained-translation} does not
work for full meet revision. Another reduction still proves it at least
\np-hard.

\state{fullmeet-hard}{Lemma}{}{

\begin{lemma}

Establishing $I \leq_{\emptyset S} J$ is \np-hard
if $S$ is a sequence of full meet revisions.

\end{lemma}

}{

\proof The proof grounds on a condition employed in the proof for natural and
severe revision: the intersection of a revision with the first class is the
formula to be established satisfiable.

\long\def\ttytex#1#2{#1}
\ttytex{
}{
+-------------+
|+---+        |
||abF| I      | a
|+-+-+        |
+--+----------+
|             |
|    J        | -a
|             |
+-------------+
}

If $F$ is consistent, full meet revision $\neg a \vee abF$ raises $abF$ and
merges the rest: $I \equiv J$. Otherwise, $abF$ is empty, and full meet
revision swaps the two clasps: $J < I$.

The claim that $I \leq_S J$ holds if and only if $F$ is satisfiable is proved
by the following models and revisions.

\begin{eqnarray*}
I	&=&	\{a\}				\\
J	&=&	\{b\}				\\
S	&=&	[a, \neg a \vee abF]
\end{eqnarray*}

All models of the first revision $a$ are minimal in the flat order:
{} $\min(a) = 0$.

\begin{eqnarray*}
\lefteqn{\emptyset \full(a)}
\\
&=& [
	\min(a),
	\true \backslash \min(a)
]
\\
&=& [
	a,
	\true \backslash a
]
\\
&=& [
	a,
	\neg a
]
\\
&=&
	C_a
\end{eqnarray*}

If $F$ is unsatisfiable, $abF$ is inconsistent. Therefore, the second revision
$S_2 = \neg a \vee abF$ is $\neg a$. Its minimal models are $\min(S_2) = \neg
a$ in the order $\emptyset \full(a) = [a, \neg a] = C_a$.

\begin{eqnarray*}
\lefteqn{C_a \full(S_2)}
\\
&=& [
	\min(S_2),
	\true \backslash \min(S_2)
]
\\
&=& [
	\neg a,
	\true \backslash \neg a
]
\\
&=& [
	\neg a,
	a
]
\end{eqnarray*}

\long\def\ttytex#1#2{#1}
\ttytex{
}{
+-------------+          +-------------+
|             |          |             |
|      I      | a        |    J        | -a
|             |          |             |
+-------------+    =>    +-------------+
|             |          |             |
|    J        | -a       |      I      |  a
|             |          |             |
+-------------+          +-------------+
}

The first model $I=\{a\}$ falsifies $\neg a$, the second $J = \{b\}$ satisfies
it: $J < I$ and $I \not\leq J$.

If $F$ is satisfiable, the minimal models of $S_2 = \neg a \vee abF$ in the
order $\emptyset \full(a) = [a, \neg a] = C_a$ are
{} $\min(S_2) = abF$.

\begin{eqnarray*}
\lefteqn{C_a \full(S_2)}
\\
&=& [
	\min(S_2),
	\true \backslash \min(S_2)
]
\\
&=& [
	abF,
	\true \backslash abF
]
\\
&=& [
	abF,
	a \wedge \neg (abF)
]
\end{eqnarray*}

\long\def\ttytex#1#2{#1}
\ttytex{
}{
                          +---+
                          |abF|
                          +-+-+
+-------------+          +-------------+
|+---+        |          |+---+        |
||abF| I      | a        ||XXX| I      |
|+-+-+        |          |+---+        | a - abF
+--+----------+    =>    |             |   u
|             |          |             | -a
|    J        | -a       |    J        |
|             |          |             |
+-------------+          +-------------+
}

Since $I$ and $J$ falsify either $a$ or $b$, they are not in the first class
$abF$. They are both in the second: $I \equiv J$ and $I \leq J$.

The comparison $I \leq J$ holds only in the second case, where $F$ is
satisfiable.~\qed

}

\draft

does not work for cone-hardness

I and J cannot be models of AB because this formula may be inconsistent

they cannot be models of -(AB) because its models are equalized in the
consistent case, no comparison I<=J is erased

placing AB below is useless because only the part above counts

if AB is added a part independent from the satisfiability of F, this part would
always minimal, and the revision would not depend on the satisfiability of F

all this hints that is not coNP-hard, and maybe in NP; but does not prove it

proof could start from evaluating F1..Fm according to I and J; but the results
are combined according to their mutual consistency, even if the result only
comprises two classes

\enddraft

\subsection{Moderate severe revision}

Moderate severe is proved only \conp-hard.

\state{moderate-severe-hard}{Lemma}{}{

\begin{lemma}

Establishing $I \leq_{\emptyset S} J$ is $\conp$-hard
if $S$ is a sequence of moderate severe revisions.

\end{lemma}

}{

\proof The proof core is a two-class order after a first revision. If the
second revision is inconsistent with the first class, it equalizes all its
non-models. This inconsistency is linked to the unsatisfiability of an
arbitrary formula $F$.

\long\def\ttytex#1#2{#1}
\ttytex{
}{
+-------------+
|+---+        |
||abF|  J     | a
|+-+-+        |
+--+----------+
|+-+---+      |
|| -ab | I    | -a
|+-----+      |
+-------------+
}

The models and revisions are:

\begin{eqnarray*}
I	&=&	\emptyset	\\
J	&=&	\{a\}		\\
S	&=&	[a, b \wedge (a \rightarrow F)]
\end{eqnarray*}

The result $\emptyset \msev(a)$ of the first revision is $C_a = [a,\neg a]$
since $\imin(a) = \imax(a) = 0$.

\begin{eqnarray*}
\lefteqn{\emptyset \msev(a)}
\\
&=& [
	\emptyset(\imin(a)) \cap a, \ldots, \emptyset(\imax(a)) \cap a,
	\emptyset(0) \cup \cdots \cup \emptyset(\imin(a)) \backslash a,
\\
&&
	\emptyset(\imin(a)+1) \backslash a, \ldots, \emptyset(m) \backslash a
]
\\
&=& [
	\emptyset(0) \cap a, \ldots, \emptyset(0) \cap a,
	\emptyset(0) \cup \cdots \cup \emptyset(0) \backslash a,
	\emptyset(1) \backslash a, \ldots, \emptyset(m) \backslash a
]
\\
&=& [
	\true \cap a,
	\true \backslash a
]
\\
&=& [
	a,
	\neg a
]
\\
&=&
	C_a
\end{eqnarray*}

The second revision $S_2 = b \wedge (a \rightarrow F)$ is equivalent to
{} $abF \vee \neg ab$.
The two disjuncts are respectively its intersection with the first and the
second class of $C_a = [a,\neg a]$. The first is non-empty if and only if $F$
is satisfiable.

If $F$ is unsatisfiable, $abF \vee \neg ab$ is equivalent to $\neg a b$.
This formula is contained in the second class $\neg a$ of the current order
$[a,\neg a]$. The relevant indexes therefore coincide:
{} $\imin(S_2) = \imax(S_2) = 1$.

\begin{eqnarray*}
\lefteqn{C_a \msev(S_2)}
\\
&=& [
	C_a(\imin(S_2)) \cap S_2, \ldots, C_a(\imax(S_2)) \cap S_2,
	C_a(0) \cup \cdots \cup C_a(\imin(S_2)) \backslash S_2,
\\
&&
	C_a(\imin(S_2)+1) \backslash S_2, \ldots, C_a(m) \backslash S_2
]
\\
&=& [
	C_a(1) \cap S_2, \ldots, C_a(1) \cap S_2,
	C_a(0) \cup \cdots \cup C_a(1) \backslash S_2,
	C_a(1+1) \backslash S_2, \ldots, C_a(1) \backslash S_2
]
\\
&=& [
	C_a(1) \cap S_2,
	C_a(0) \cup C_a(1) \backslash S_2
]
\\
&=& [
	\neg a \wedge \neg a b,
	\true \wedge \neg (\neg a b)
]
\\
&=& [
	\neg a b,
	a \vee \neg b
]
\end{eqnarray*}

\long\def\ttytex#1#2{#1}
\ttytex{
}{
                          +-----+
                          | -ab |
                          +-----+
+-------------+          +-------------+
|             |          |             |
|       J     | a        |       J     |
|             |          |             |
+-------------+    =>    |             | a v -b
|+-----+      |          |+-----+      |
|| -ab | I    | -a       ||XXXXX| I    |
|+-----+      |          |+-----+      |
+-------------+          +-------------+
}

Both models $I$ and $J$ falsify $\neg a \wedge b$ and are therefore in the
second class. As a result, $I \leq J$.

If $F$ is satisfiable, the intersection of the second revision
{} $S_2 = b \wedge (a \rightarrow F)$
with the first class of $C = [a,\neg a]$ is $abF$, which is not empty. Neither
is the intersection $a \neg b$ with the second class. The conclusions
{} $\imin(S_2) = 0$
and
{} $\imin(S_2) = 1$
allow calculating the final order.

\begin{eqnarray*}
\lefteqn{C_a \msev(S_2)}
\\
&=& [
	C_a(\imin(S_2)) \cap S_2, \ldots, C_a(\imax(S_2)) \cap S_2,
\\
&&
	C_a(0) \cup \cdots \cup C_a(\imin(S_2)) \backslash S_2,
	C_a(\imin(S_2)+1) \backslash S_2, \ldots, C_a(m) \backslash S_2
]
\\
&=& [
	C_a(0) \cap S_2, \ldots, C_a(1) \cap S_2,
	C_a(0) \cup \cdots \cup C_a(0) \backslash S_2,
	C_a(0+1) \backslash S_2, \ldots, C_a(1) \backslash S_2
]
\\
&=& [
	C_a(0) \cap S_2,
	C_a(1) \cap S_2,
	C_a(0) \backslash S_2,
	C_a(1) \backslash S_2
]
\\
&=& [
	a \cap S_2,
	\neg a \cap S_2,
	a \backslash S_2,
	\neg a \backslash S_2
]
\end{eqnarray*}

\long\def\ttytex#1#2{#1}
\ttytex{
}{
                          +---+
                          |abF|
                          +-+-+
                          +-+---+
                          | -ab |
                          +-----+
+-------------+          +-------------+
|+---+        |          |+---+        |
||abF|  J     | a        ||XXX|  J     | a - abF
|+-+-+        |          |+-+-+        |
+--+----------+    =>    +--+----------+
|+-+---+      |          |+-+---+      |
|| -ab | I    | -a       ||XXXXX| I    | -a -(-a b)
|+-----+      |          |+-----+      |
+-------------+          +-------------+
}

Both $I$ and $J$ falsify $b$ are therefore $S_2$: none of them is in the first
or second class of this order. The first model $I = \emptyset$ falsifies $a$
and is therefore not in the third class. The second model $J = \{a\}$ satisfies
$a$ and is therefore in the third class. The conclusion $J < I$ proves the
claim $I \not\leq J$.~\qed

}

\draft

neither I nor J can be in F because this formula may be inconsistent; they
cannot be models of the rest of the revision because these models always
maintain their mutual order and always go above the other, regardless of F

the only hook for hardness is the possible equalization of models that do not
satisfy the revision

this is why is only coNP-hard

\enddraft

\subsection{Deep severe revision}

Deep severe revision is \np-hard, like full meet revision.

\state{deep-severe-hard}{Lemma}{}{

\begin{lemma}
\statelabel

Establishing $I \leq_{\emptyset S} J$ is $\np$-hard
if $S$ is a sequence of deep severe revisions.

\end{lemma}

}{

\proof The claim is proved by the following models and sequence of revisions.

\begin{eqnarray*}
I	&=&	\emptyset	\\
J	&=&	\{a\}		\\
S	&=&	[a, b \wedge (\neg a \rightarrow F)]
\end{eqnarray*}

The root of the proof is the order $C_a = [a,\neg a]$ resulting from revising
the flat order $\emptyset$ by $a$.

\long\def\ttytex#1#2{#1}
\ttytex{
}{
+-------------+
|+-----+      |
|| ab  |  J   | a
|+--+--+      |
+---+---------+
|+--+--+      |
||-abF |  I   | -a
|+-----+      |
+-------------+
}

Te second revision $S_2 = b \wedge (\neg a \rightarrow F)$ is equivalent to
$ab \vee \neg abF$, which is part in the first class and part of the second.
Deep severe revision lifts both parts and equates the other models of the same
classes. The second part $\neg abF$ only exists if $F$ is consistent. Deep
severe revision equates $I$ with $J$ only in this case.

The result $\emptyset \dsev(a)$ of the first revision is $C_a = [a,\neg a]$
since $\imax(a) = 0$.

\begin{eqnarray*}
\lefteqn{\emptyset \dsev(a)}
&
\\
&=&
	\emptyset \dsev(a)
\\
&=&
[
	\emptyset(0) \cap a, \ldots, \emptyset(m) \cap a,
	\emptyset(0) \cup \cdots \cup \emptyset(\imax(a)) \backslash a, \\
&&
	\emptyset(\imax(a) + 1), \ldots, \emptyset(m)
] 
\\
&=&
[
	\emptyset(0) \cap a, \ldots, \emptyset(0) \cap a,
	\emptyset(0) \cup \cdots \cup \emptyset(0) \backslash a, \\
&&
	\emptyset(0 + 1), \ldots, \emptyset(m)
] 
\\
&=&
[
	\emptyset(0) \cap a,
	\emptyset(0) \backslash a
]
\\
&=&
[
	\true \wedge a,
	\true \wedge \neg a
]
\\
&=&
[
	a,
	\neg a
]
\\
&=&
	C_a
\end{eqnarray*}

If $F$ is inconsistent, the second revision
{} $S_2 = b \wedge (\neg a \rightarrow F)$
is equivalent to
{} $b \wedge (a \vee \false)$ = $a \wedge b$.
It is consistent with the first class $a$ of $C_a$ and inconsistent with the
second $\neg a$. As a result, its maximal index is zero:
{} $\imax(S_2) = 1$.

\begin{eqnarray*}
\lefteqn{C_a \dsev(S_2)}
\\
&=&
[
	C_a(0) \cap S_2, \ldots, C_a(m) \cap S_2,
	C_a(0) \cup \cdots \cup C_a(\imax(S_2)) \backslash S_2, \\
&&
	C_a(\imax(S_2) + 1), \ldots, C_a(m)
] 
\\
&=&
[
	C_a(0) \cap S_2, C_a(1) \cap S_2,
	C_a(0) \backslash S_2,
	C_a(0 + 1), C_a(1)
] 
\\
&=&
[
	S_2, \emptyset,
	a \backslash S_2,
	C_a(1)
] 
\\
&=&
[
	a \wedge b,
	a \wedge \neg (a \wedge b),
	\neg a
] 
\\
&=&
[
	a \wedge b,
	a \wedge \neg b,
	\neg a
] 
\end{eqnarray*}

\long\def\ttytex#1#2{#1}
\ttytex{
}{
                          +-----+
                          | ab  |
                          +-----+
+-------------+          +-------------+
|+-----+      |          |+-----+      |
|| ab  |  J   | a        ||XXXXX|  J   | a - ab = a-b
|+-----+      |          |+-----+      |
+-------------+    =>    +-------------+
|             |          |             |
|         I   | -a       |         I   | -a
|             |          |             |
+-------------+          +-------------+
}

The second class $a \wedge \neg b$ contains $J = \{a\}$, the third $\neg a$
contains $I = \emptyset$. Therefore, $J < I$.

If $F$ is consistent, the second revision $S_2$ is consistent with the second
class $\neg a$ of $C_a$. Indeed, $b \wedge (\neg a \rightarrow F) \wedge \neg
a$ is equivalent to $\neg a b F$, which is satisfiable like $F$. The maximum
index of $S_2$ is therefore one: $\imax(S_2) = 1$.

\begin{eqnarray*}
\lefteqn{C_a \dsev(S_2)}
\\
&=&
[
	C_a(0) \cap S_2, \ldots, C_a(m) \cap S_2,
	C_a(0) \cup \cdots \cup C_a(\imax(S_2)) \backslash S_2, \\
&&
	C_a(\imax(S_2) + 1), \ldots, C_a(m)
] 
\\
&=&
[
	C_a(0) \cap S_2, \ldots, C_a(1) \cap S_2,
	C_a(0) \cup \cdots \cup C_a(1) \backslash S_2, \\
&&
	C_a(1 + 1), \ldots, C_a(1)
] 
\\
&=&
[
	C_a(0) \cap S_2, C_a(1) \cap S_2,
	C_a(0) \cup C_a(1) \backslash S_2
] 
\\
&=&
[
	a \cap S_2, \neg a \cap S_2,
	\true \backslash S_2
] 
\end{eqnarray*}

\long\def\ttytex#1#2{#1}
\ttytex{
}{
                          +-----+
                          | ab  |
                          +-----+
                          +-----+
                          |-abF |
                          +-----+
+-------------+          +-------------+
|+-----+      |          |+-----+      |
|| ab  |  J   | a        ||XXXXX|  J   |
|+--+--+      |          |+-----+      |
+---+---------+    =>    |             |
|+--+--+      |          |+-----+      |
||-abF |  I   | -a       ||XXXXX|  I   |
|+-----+      |          |+-----+      |
+-------------+          +-------------+
}

Neither $I$ nor $J$ satisfy $S_2 = b \wedge (\neg a \rightarrow F)$ since they
both evaluate $b$ to false. As a result, they are both in the third class. They
are equivalent: $I \equiv J$. Contrarily to the previous case, $I \leq J$
holds.~\qed

}

\subsection{Lexicographic and very radical revisions}

These revisions do not involve either on the minimum and maximal models of the
revision. This saves them from a check of existence: no model is less than a
minimal one, and none is greater than a maximal. This makes sorting models
easy.

The definition of radical revision includes the minimum and maximal indexes of
the revision, but an equivalent formulation does not:
{} $C \vrad(A) = [C(0) \cap A, \ldots, C(m) \cap A, \true \backslash A ]$.

\state{lexicographic-veryradical-polynomial}{Lemma}{}{

\begin{lemma}
\statelabel

Establishing $I \leq_{C S} J$ and $J \leq_{C S}$ is polynomial if $S$ is a
possibly heterogeneous sequence of lexicographic and very radical revisions,
and $I \leq_C J$ and $J \leq_C I$ are polynomial.

\end{lemma}

}{

\proof The claim is proved inductively: $I <_{C \rev(A)} J$ is polynomially
reduced to the comparison $I <_C J$, and the same for equivalence.

\begin{description}

\item[Very radical revision:]

\[
C \vrad(A) = [
	C(\imin(A)) \cap A, \ldots, C(\imax(A)) \cap A,
	\true \backslash A
]
\]

Two models $I,J$ are equivalent in $C \vrad(A)$ if they are in the same class
of
{} $[C(\imin(A)) \cap A, \ldots, C(\imax(A)) \cap A, \true \backslash A]$:
either $I,J \in C(i) \cap A$ or $I,J \in \true \backslash A$. Either $I,J \in
C(i)$ and $I,J \in A$ or $I,J \not\in A$. Membership to the same class $C(i)$
is equivalence in $C$.

A model $I$ is strictly less in $C \vrad(A)$ than all others $J$ belonging to
classes of higher index in
{} $[C(\imin(A)) \cap A, \ldots, C(\imax(A)) \cap A, \true \backslash A]$:
either $I \in C(i) \cap A$ and $J \in C(j) \cap A$ with $i<j$, or $I \in A$ and
$J \not\in A$. The first alternative is the same as
{} $I \in C(i)$, $J \in C(j)$ and $I,J \in A$, with $i < j$.
Strict comparison in $C$ is
{} $I \in C(i)$, $J \in C(j)$ and $i < j$.

\item[Lexicographic revision:]

\[
C \lex(A) = [
	C(0) \cap A, \ldots, C(m) \cap A,
	C(0) \backslash A, \ldots, C(m) \backslash A
]
\]

Two models $I,J$ are equivalent in $C \lex(A)$ if they are in the same class of
{} $C = [C(0) \cap A, \ldots, C(m) \cap A,
{}       C(0) \backslash A, \ldots, C(m) \backslash A]$:
either $I,J \in C(i) \cap A$ or $I,J \in C(i) \backslash A$. The common
subcondition $I,J \in C(i)$ is equivalence in $C$.

A model $I$ is strictly less in $C \lex(A)$ than all others $J$ belonging to
classes of higher index in
{} $C = [C(0) \cap A, \ldots, C(m) \cap A,
{}       C(0) \backslash A, \ldots, C(m) \backslash A]$:
either $I \in C(i) \cap A$ and $J \in C(j) \cap A$ with $i<j$,
{} or
either $I \in C(i) \backslash A$ and $J \in C(j) \backslash A$ with $i<j$
{} or
$I \in A$ and $J \not\in A$. The first two alternatives include
$I \in C(i)$, $J \in C(j)$ with $i < j$, which is strict
comparison in $C$.

Overall, both comparisons of both revisions are expressed in terms of
evaluation of $A$ in both models and the same evaluation of the same models in
$C$. Evaluation of $A$ is polynomial. Induction proves polynomiality.

Non-strict comparison is either equivalence or strict comparison, both
polynomial.

\end{description}
\qed

}

\draft
\long\def\ttytex#1#2{#1}
\ttytex{
}{

C vrad(A) = [C(imin(A))A .. C(imax(A))A true-A]

I==J holds if I,JcA e I==CJ or I,J/cA
two evaluations IcA,JcA and the inductive test I==CJ

I<J holds if IcA,J/cA or I,JcA and I<CJ
two evaluations IcA,JcA and the inductive text I<CJ

C lex(A) = [C(0)A .. C(m)A C(0)-A .. C(m)-A]

I==J holds if I,JcA e I==CJ or I,J/cA e I==CJ
two evaluations and the inductive text

I<J holds if IcA,J/cA or I,JcA e I<CJ or I,J/cA e I<CJ
two evaluations and the inductive text

for all: every pair of comparisons in Crev(A) requires the test IcA,JcA given
THE SAME comparisons in C

inductively, it is polynomial

}
\enddraft

\draft

\subsection{construction of models and formulae}

example: Dp-hardness of natural revision

\v
FcA
Gc-A
JcA-F
Ic-A-G
\vv

\long\def\ttytex#1#2{#1}
\ttytex{
}{
+-------------+
|+---+        |
|| F | J      | A
|+---+        |
+-------------+
|        +---+|
|      I | G || -A
|        +---+|
+-------------+
}

four models are needed, forgetting the variables X of F and Y of G

A is variable a

the difference between F/J and I/G is variable b

\long\def\ttytex#1#2{#1}
\ttytex{
}{
   b      -b
+-------------+
|+---+        |
|| F |     J  | a
|+---+        |
+-------------+
|        +---+|
|  I     | G || -a
|        +---+|
+-------------+
}

both I and J have X=0, Y=0

F and G have arbitrary values for the other variables

\v
I =  -a b 0 0
J =  a -b 0 0
F =   a b F *
G = -a -b * G
\vv

revisions: [A,-AuF,AuG]

\v
A = a
-AuF = -a u abF
AuG =  a u -a-bG
\vv

\enddraft

\section{Necessary and equivalent conditions}
\label{condition}

This section and the following concern lexicographic revision only, allowing
for the notational simplification of suppressing the kind of revision:

\begin{itemize}

\item $\emptyset A$ stands for $\emptyset \lex(A)$;

\item $\emptyset [S_1,\ldots,S_m]$
stands for
$\emptyset [\lex(S_1),\ldots,\lex(S_m)]$.

\end{itemize}

\subsection{The expression of redundancy}

The redundancy of $S1$ in $[S_1,S_2,\ldots,S_m]$ from the flat doxastic state
$\emptyset$ is proved the same as the equivalence of $S_1$ with a disjunction
of conjunctions of all formulae $S_2,\ldots,S_m$, each possibly negated.

For example, the redundancy of $S_1$ in $[S_1,S_2,S_3]$ from the flat doxastic
state $\emptyset$ is the equivalence of $S_1$ with either
{} $S_2 \wedge S_3$,
{} or to $S_2 \wedge \neg S_3$,
{} or to $\neg S_2 \wedge S_3$,
{} or to $\neg S_2 \wedge \neg S_3$,
{} or to a disjunction of some of these formulae.

Formally, redundancy is the equivalence of $S_1$ with the disjunction of some
formulae
{} $Q = (B_2 \equiv S_2) \wedge \cdots \wedge (B_m \equiv S_m)$,
where each $B_i$ is either $\true$ or $\false$.

\subsection{Overview}

This expression of redundancy has an intuitive explanation. Like the others,
the oldest revision $S_1$ only states that some models are plausible and the
others implausible. It is irrelevant if the other revisions $S_2,\ldots,S_m$
also distinguish them by sorting them the same or the opposite: the other
revisions either replicate or overrule the first.

The oldest revision $S_1$ sets the models satisfying the revising formula lower
in the preorder than the others. It partitions the set of all models in two
groups, as depicted in Figure~\ref{groups}.

\begin{hfigure}
\setlength{\unitlength}{2500sp}%
\begin{picture}(3066,1866)(2518,-3244)
\thicklines
{\color[rgb]{0,0,0}\put(2551,-3211){\framebox(3000,1800){}}
}%
{\color[rgb]{0,0,0}\put(2551,-2311){\line( 1, 0){3000}}
}%
\put(4051,-1936){\makebox(0,0)[b]{\smash{\fontsize{6}{7.2}
\usefont{T1}{cmr}{m}{n}{\color[rgb]{0,0,0}$0$}%
}}}
\put(4051,-2836){\makebox(0,0)[b]{\smash{\fontsize{6}{7.2}
\usefont{T1}{cmr}{m}{n}{\color[rgb]{0,0,0}$1$}%
}}}
\end{picture}%
\nop{
+-----------------------+
|                       |
|          0            |
|                       |
+-----------------------+
|                       |
|          1            |
|                       |
+-----------------------+
}
\label{groups}
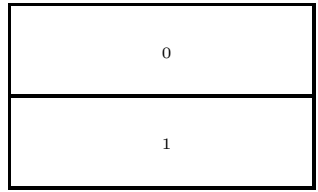
\hcaption{groups of $S_1$}
\end{hfigure}

The other revisions $S_2,\ldots,S_m$ separate the models their way. They may
refine the separation of $S_1$; they do if their boundaries include the
boundary of $S_1$: models of different groups of $S_1$ are in different groups
of $S_2,\ldots,S_m$. The order between the groups does not matter. The models
in their group 3 may compare less than the models in their group 2 when they
compare greater according to $S_1$. What counts are the groups themselves, not
their order. In the figure, the lines and not the numbers.

\begin{hfigure}
\begin{tabular}{ccc}
\setlength{\unitlength}{2500sp}%
\begin{picture}(3066,1866)(2518,-3244)
\thicklines
{\color[rgb]{0,0,0}\put(2551,-3211){\framebox(3000,1800){}}
}%
{\color[rgb]{0,0,0}\put(2551,-2311){\line( 1, 0){3000}}
}%
\put(4051,-1936){\makebox(0,0)[b]{\smash{\fontsize{6}{7.2}
\usefont{T1}{cmr}{m}{n}{\color[rgb]{0,0,0}$0$}%
}}}
\put(4051,-2836){\makebox(0,0)[b]{\smash{\fontsize{6}{7.2}
\usefont{T1}{cmr}{m}{n}{\color[rgb]{0,0,0}$1$}%
}}}
\end{picture}%
\nop{
+-----------------------+
|                       |
|          0            |
|                       |
+-----------------------+
|                       |
|          1            |
|                       |
+-----------------------+
}
& ~~~ &
\setlength{\unitlength}{2500sp}%
\begin{picture}(3066,1866)(2518,-3244)
\thicklines
{\color[rgb]{0,0,0}\put(2551,-3211){\framebox(3000,1800){}}
}%
{\color[rgb]{0,0,0}\multiput(3001,-2311)(12.00000,-12.00000){26}{\makebox(13.3333,20.0000){\small.}}
\multiput(3301,-2611)(-12.09677,-12.09677){32}{\makebox(13.3333,20.0000){\small.}}
\multiput(2926,-2986)(14.42308,-8.65385){27}{\makebox(13.3333,20.0000){\small.}}
}%
{\color[rgb]{0,0,0}\put(3151,-2761){\line( 3,-2){450}}
\put(3601,-3061){\line( 2, 3){507.692}}
}%
{\color[rgb]{0,0,0}\put(3826,-2761){\line( 1, 0){525}}
\multiput(4351,-2761)(10.71429,-12.85714){36}{\makebox(13.3333,20.0000){\small.}}
}%
{\color[rgb]{0,0,0}\put(4501,-2911){\line( 3, 4){450}}
}%
{\color[rgb]{0,0,0}\put(3301,-2311){\line( 0, 1){600}}
\put(3301,-1711){\line( 3, 2){450}}
}%
{\color[rgb]{0,0,0}\put(3751,-2161){\line( 1, 1){450}}
\put(4201,-1711){\line( 1, 0){450}}
\put(4651,-1711){\line( 1,-1){600}}
}%
{\color[rgb]{0,0,0}\multiput(4951,-2011)(7.50000,15.00000){41}{\makebox(13.3333,20.0000){\small.}}
}%
{\color[rgb]{0,0,0}\put(3301,-1711){\line( 1,-1){600}}
}%
{\color[rgb]{0,0,0}\put(2551,-2311){\line( 1, 0){3000}}
}%
\put(2851,-2761){\makebox(0,0)[b]{\smash{\fontsize{6}{7.2}
\usefont{T1}{cmr}{m}{n}{\color[rgb]{0,0,0}$2$}%
}}}
\put(3526,-2611){\makebox(0,0)[b]{\smash{\fontsize{6}{7.2}
\usefont{T1}{cmr}{m}{n}{\color[rgb]{0,0,0}$5$}%
}}}
\put(3451,-2161){\makebox(0,0)[b]{\smash{\fontsize{6}{7.2}
\usefont{T1}{cmr}{m}{n}{\color[rgb]{0,0,0}$1$}%
}}}
\put(2851,-1861){\makebox(0,0)[b]{\smash{\fontsize{6}{7.2}
\usefont{T1}{cmr}{m}{n}{\color[rgb]{0,0,0}$3$}%
}}}
\put(3751,-1861){\makebox(0,0)[b]{\smash{\fontsize{6}{7.2}
\usefont{T1}{cmr}{m}{n}{\color[rgb]{0,0,0}$7$}%
}}}
\put(5326,-1936){\makebox(0,0)[b]{\smash{\fontsize{6}{7.2}
\usefont{T1}{cmr}{m}{n}{\color[rgb]{0,0,0}$8$}%
}}}
\put(4426,-2086){\makebox(0,0)[b]{\smash{\fontsize{6}{7.2}
\usefont{T1}{cmr}{m}{n}{\color[rgb]{0,0,0}$0$}%
}}}
\put(4351,-2611){\makebox(0,0)[b]{\smash{\fontsize{6}{7.2}
\usefont{T1}{cmr}{m}{n}{\color[rgb]{0,0,0}$9$}%
}}}
\put(5176,-2836){\makebox(0,0)[b]{\smash{\fontsize{6}{7.2}
\usefont{T1}{cmr}{m}{n}{\color[rgb]{0,0,0}$4$}%
}}}
\put(3976,-3061){\makebox(0,0)[b]{\smash{\fontsize{6}{7.2}
\usefont{T1}{cmr}{m}{n}{\color[rgb]{0,0,0}$6$}%
}}}
\end{picture}%
\nop{
+----------+---+--------+
|  3       |   |        |
|      +---+   |     5  |
|      |   1   |        |
+------+----+--+--------+
|    2      |    6      |
|   +-------+----+------+
|   |     0      |  4   |
+---+------------+------+
}
\end{tabular}
\label{refine}
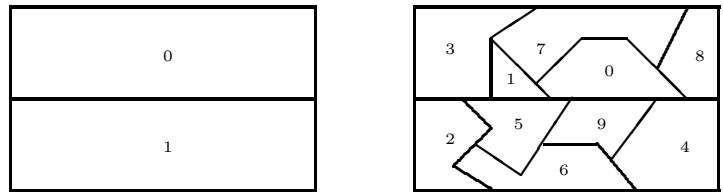
\hcaption{$S_2,\ldots,S_m$ refine the groups of $S_1$}
\end{hfigure}

The groups 0 and 1 of $S_1$ are each the union of some groups of
{} $S_2,\ldots,S_{m-1}$
in Figure~\ref{refine}. Two models of different groups of $S_1$ are in
different groups of
{} $S_2,\ldots,S_m$, which
decide their order because they take precedence over the previous $S_1$ in
lexicographic revision. Whether $S_1$ sorts models the same or the opposite
does not matter.

If the other revisions refine the order of the oldest they make it redundant.
Otherwise, the other revisions
{} $S_2,\ldots,S_m$
do not refine the separation of $S_1$: some of their groups include models that
are in different groups of $S_1$. In Figure~\ref{no-refine}, the models in
their group 2 may include models in both groups 0 and 1 of $S_1$. The other
revisions
{} $S_2,\ldots,S_m$
compare them the same. The oldest revision $S_1$ decides. It matters. It is not
redundant.

\begin{hfigure}
\begin{tabular}{ccc}
\setlength{\unitlength}{2500sp}%
\begin{picture}(3066,1866)(2518,-3244)
\thicklines
{\color[rgb]{0,0,0}\put(2551,-3211){\framebox(3000,1800){}}
}%
{\color[rgb]{0,0,0}\put(2551,-2311){\line( 1, 0){3000}}
}%
\put(4051,-1936){\makebox(0,0)[b]{\smash{\fontsize{6}{7.2}
\usefont{T1}{cmr}{m}{n}{\color[rgb]{0,0,0}$0$}%
}}}
\put(4051,-2836){\makebox(0,0)[b]{\smash{\fontsize{6}{7.2}
\usefont{T1}{cmr}{m}{n}{\color[rgb]{0,0,0}$1$}%
}}}
\end{picture}%
\nop{
+-----------------------+
|                       |
|          0            |
|                       |
+-----------------------+
|                       |
|          1            |
|                       |
+-----------------------+
}
& ~~~ &
\setlength{\unitlength}{2500sp}%
\begin{picture}(3066,1866)(2518,-3244)
\thicklines
{\color[rgb]{0,0,0}\put(2551,-3211){\framebox(3000,1800){}}
}%
{\color[rgb]{0,0,0}\put(3001,-2311){\line( 1,-1){450}}
\put(3451,-2761){\line( 0,-1){450}}
}%
{\color[rgb]{0,0,0}\put(3301,-2611){\line( 1, 0){900}}
}%
{\color[rgb]{0,0,0}\put(4051,-2761){\line( 1,-1){450}}
}%
{\color[rgb]{0,0,0}\put(4951,-2311){\line( 0,-1){900}}
}%
{\color[rgb]{0,0,0}\put(4051,-2761){\line( 1, 1){750}}
\put(4801,-2011){\line(-1, 0){1500}}
}%
{\color[rgb]{0,0,0}\put(4501,-2311){\line( 1, 0){1050}}
}%
{\color[rgb]{0,0,0}\put(2551,-2311){\line( 1, 0){450}}
\put(3001,-2311){\line( 1, 1){450}}
\put(3451,-1861){\line(-1, 1){450}}
}%
{\color[rgb]{0,0,0}\put(4051,-2011){\line( 0, 1){600}}
}%
{\color[rgb]{0,0,0}\put(4501,-2011){\line( 1, 1){600}}
}%
\thinlines
{\color[rgb]{0,0,0}\put(3901,-2311){\line( 1, 0){150}}
}%
{\color[rgb]{0,0,0}\put(3526,-2311){\line( 1, 0){150}}
}%
{\color[rgb]{0,0,0}\put(3151,-2311){\line( 1, 0){150}}
}%
{\color[rgb]{0,0,0}\put(4276,-2311){\line( 1, 0){150}}
}%
\put(4426,-1786){\makebox(0,0)[b]{\smash{\fontsize{6}{7.2}
\usefont{T1}{cmr}{m}{n}{\color[rgb]{0,0,0}$0$}%
}}}
\put(3676,-1786){\makebox(0,0)[b]{\smash{\fontsize{6}{7.2}
\usefont{T1}{cmr}{m}{n}{\color[rgb]{0,0,0}$1$}%
}}}
\put(4651,-2911){\makebox(0,0)[b]{\smash{\fontsize{6}{7.2}
\usefont{T1}{cmr}{m}{n}{\color[rgb]{0,0,0}$4$}%
}}}
\put(5176,-2011){\makebox(0,0)[b]{\smash{\fontsize{6}{7.2}
\usefont{T1}{cmr}{m}{n}{\color[rgb]{0,0,0}$6$}%
}}}
\put(3826,-2986){\makebox(0,0)[b]{\smash{\fontsize{6}{7.2}
\usefont{T1}{cmr}{m}{n}{\color[rgb]{0,0,0}$8$}%
}}}
\put(2926,-2911){\makebox(0,0)[b]{\smash{\fontsize{6}{7.2}
\usefont{T1}{cmr}{m}{n}{\color[rgb]{0,0,0}$7$}%
}}}
\put(2851,-2011){\makebox(0,0)[b]{\smash{\fontsize{6}{7.2}
\usefont{T1}{cmr}{m}{n}{\color[rgb]{0,0,0}$3$}%
}}}
\put(5251,-2836){\makebox(0,0)[b]{\smash{\fontsize{6}{7.2}
\usefont{T1}{cmr}{m}{n}{\color[rgb]{0,0,0}$5$}%
}}}
\put(3751,-2386){\makebox(0,0)[b]{\smash{\fontsize{6}{7.2}
\usefont{T1}{cmr}{m}{n}{\color[rgb]{0,0,0}$2$}%
}}}
\end{picture}%
\nop{
+-----------------------+
|   4  +------+----+----|
|      |      |    | 5  |
+------+   3  |    |    |
+-------+.....|..2.+----+
|       |     |         |
|   1   +-----+-----+   |
|       |     0     |   |
+-------+-----------+---+
}
\end{tabular}
\label{no-refine}
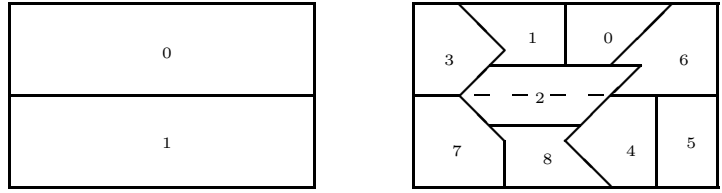
\hcaption{$S_2,\ldots,S_m$ do not refine the groups of $S_1$}
\end{hfigure}

The groups of $S_2,\ldots,S_m$ redefine redundancy. The revisions
{} $S_2,\ldots,S_m$
separate the models like $S_1$ does: $S_2$ and $\neg S_2$, $S_3$ and $\neg S_3$
and so on. Two models are in the same group only if they evaluate every formula
$S_i$ the same. If they evaluate a single formula $S_i$ differently, they are
in different groups. If they evaluate no formula $S_i$ differently, they are in
the same group. As a result, a formula like
{} $S_1 \wedge \neg S_2 \wedge \cdots \wedge \neg S_m$
bounds one group, for example. Two of its models evaluate every formulae the
same, and are therefore in the same group. They evaluate $S_1$ to true while
the models of another formula
{} $\neg S_1 \wedge \neg S_2 \wedge \cdots \wedge \neg S_m$,
evaluate $S_1$ to false, and are therefore in a different group.

These conjunctions are called Q-combinations. The first revision is redundant
if and only if its two groups each comprise the union of groups of some
Q-combinations.

\draft

A concrete example is the sequence of lexicographic revisions
$[S_1,S_2,S_3,S_4,S_5]$ comprising the following formulae.

\begin{eqnarray*}
S_1	&=&	y \wedge (x \equiv z)					\\
S_2	&=&	\neg x \wedge (\neg y \vee \neg z)			\\
S_3	&=&	y \wedge (x \vee z)					\\
S_4	&=&	y \wedge (\neg x \vee \neg z)				\\
S_5	&=&	x \wedge (\neg y \vee z)
\end{eqnarray*}

The first revision $S_1 = y \wedge (x \equiv z)$ supports its models
$\{x,y,z\}$ and $\{\neg x, y, \neg z\}$ and negates the others. For example, it
supports $\{x,y,z\}$ over its non-model $\{\neg x, \neg y, \neg z\}$. The other
revisions change this order. If they do not distinguish $\{x,y,z\}$ and $\{\neg
x, \neg y, \neg z\}$, still $S_1$ does, breaking the tie. This is however not
the case for this example: the other revisions $[S_2,S_3,S_4,S_5]$ alone
separate the models of $S_1$ from the others. They do not support them like
$S_1$ does, they just support them differently.

\begin{eqnarray*}
S_3 \wedge S_5
&=&
	y \wedge (x \vee z)
		\wedge
	x \wedge (\neg y \vee z)
\\
&\equiv&
	y \wedge (\true \vee z)
		\wedge
	x \wedge (\neg \true \vee z)
\\
&\equiv&
	y \wedge \true
		\wedge
	x \wedge (\false \vee z)
\\
&\equiv&
	x \wedge y \wedge z
\\
S_2 \wedge S_4
&=&
	\neg x \wedge (\neg y \vee \neg z)
		\wedge
	y \wedge (\neg x \vee \neg z)
\\
&\equiv&
	\neg x \wedge (\neg \true \vee \neg z)
		\wedge
	y \wedge (\neg \false \vee \neg z)
\\
&\equiv&
	\neg x \wedge (\false \vee \neg z)
		\wedge
	y \wedge (\true \vee \neg z)
\\
&\equiv&
	\neg x \wedge \neg z
		\wedge
	y \wedge \true
\\
&\equiv&
	\neg x \wedge y \wedge \neg z
\end{eqnarray*}

All simplifications are due to unit propagation: unit clauses like $\neg x$
turn all other instances of variable $x$ into $\false$.

The first model $\{x,y,z\}$ of $S_1$ is the only model of $S_5 \wedge S_3$ and
falsifies $S_4$ and $S_2$. Therefore, it is the only model of $S_5 \wedge \neg
S_4 \wedge S_3 \wedge \neg S_2$. The second model $\{\neg x, y, \neg z\}$ is
the only model of $S_4 \wedge S_2$ and falsifies $S_5$ and $S_3$. Therefore, it
is the only model of $\neg S_5 \wedge S_4 \wedge \neg S_3 \wedge S_2$.

The order $\emptyset [S_2,S_3,S_4,S_5]$ do not have a single class for the two
models $\{x,y,z\}$ and $\{\neg x, y, \neg z\}$ of $S_1$. Yet, one of their
classes comprises the first and another the second. They support each
differently from all others. Arbitration by $S_1$ is unnecessary. It is
redundant.

\enddraft

\subsection{Equivalence between models}

The definition of the lexicographic revision has an alternative formulation
based on how a single formula orders the models.

\begin{definition}
\label{formula-order}

The order of a single formula $A$ compares $I \leq_A J$ if and only if either
$I \in A$ or $J \not\in A$.

\end{definition}

The inductive formulation of an the order produced by a sequence of
lexicographic revisions hinges around a single lexicographic revision.

\state{lexicographic-inductive}{Theorem}{}{

\begin{theorem}

For every formula $A$ and doxastic state $C$, it holds

\[
I \leq_{CA} J 	\IFF	I \leq_A J \AND (J \not\leq_A I \OR I \leq_C J).
\]

\end{theorem}

}{

\proof The class of a model $I$ in the order $C$ is denoted $C(I)$. This is the
unique number $i$ such that $I \in C(i)$. It is unique since $C$ is a
partition. By definition, $I \leq_C I$ is the same as $C(I) \leq C(J)$.

The claim is that $I \leq_{CA} J$ is equivalent to
{} $I \leq_A J \AND (J \not\leq_A I \OR I \leq_C J)$.
The definition of $I \leq_{CA} J$ is that the class index of $I$ in $CA$ is less
than or equal to the class index of $J$. The classes are
{} $CA =
{}  [C(0) \cap A, \ldots C(m) \cap A,
{}   C(0) \backslash A, \ldots C(m) \backslash A]$.
Therefore,
{} if $I \in A$ then $CA(I) = C(I)$ and
{} if $I \not\in A$ then $CA(I) = m + 1 + C(I)$.

The two conditions
{} $I \leq_{CA} J$
and
{} $I \leq_A J \AND (J \not\leq_A I \OR I \leq_C J)$
are evaluated in the four cases separated by the membership of $I$ and $J$ to
$A$.

\begin{description}

\item[$I \in A$,     $J \in A$]

\begin{eqnarray*}
I \in A &\Rightarrow& CA(I) = C(I)
\\
J \in A &\Rightarrow& CA(J) = C(J)
\\
I \leq_{CA} J
\\
&\Leftrightarrow&
                Clex(A)(I) \leq Clex(A)(J)
\\
&\Leftrightarrow&
                C(I) \leq C(J)
\\
&\Leftrightarrow&
                I \leq_C J
\\
\\
I \leq_A J
\\
&\Leftrightarrow&
                I \in A \OR J \not\in A
\\
&\Leftrightarrow&
                \true \OR \false
\\
&\Leftrightarrow&
                \true
\\
J \leq_A I
\\
&\Leftrightarrow&
                J \in A \OR I \not\in A
\\
&\Leftrightarrow&
                \true \OR \false
\\
&\Leftrightarrow&
                \true
\\
I \leq_A J \AND (J \not\leq_A I \OR I \leq_C J)
\\
&\Leftrightarrow&
                I \leq_A J \AND (\NOT J \leq_A I \OR I \leq_C J)
\\
&\Leftrightarrow&
                \true \AND (\NOT true \OR I \leq_C J)
\\
&\Leftrightarrow&
                \true \AND (\false \OR I \leq_C J)
\\
&\Leftrightarrow&
                \false \OR I \leq_C J
\\
&\Leftrightarrow&
                I \leq_C J
\end{eqnarray*}

\item[$I \in A$,     $J \not\in A$]

\begin{eqnarray*}
I \in A &\Rightarrow& CA(I) = C(I)
\\
J \not\in A &\Rightarrow& CA(J) = m + 1 + C(J)
\\
&&
I \leq_{CA} J
\\
&\Leftrightarrow&
                Clex(A)(I) \leq Clex(A)(J)
\\
&\Leftrightarrow&
                C(I) <= m + 1 + C(J)
\\
&\Leftrightarrow&
                \true
\\
\\
I \leq_A J
\\
&\Leftrightarrow&
                I \in A \OR J \not\in A
\\
&\Leftrightarrow&
                \true \OR \true
\\
&\Leftrightarrow&
                \true
\\
J \leq_A I
\\
&\Leftrightarrow&
                J \in A \OR I \not\in A
\\
&\Leftrightarrow&
                \false \OR \false
\\
&\Leftrightarrow&
                \false
\\
I \leq_A J \AND (J \not\leq_A I \OR I \leq_C J)
\\
&\Leftrightarrow&
                I \leq_A J \AND (\NOT J \leq_A I \OR I \leq_C J)
\\
&\Leftrightarrow&
                \true \AND (\NOT \false \OR I \leq_C J)
\\
&\Leftrightarrow&
                \NOT \false \OR I \leq_C J
\\
&\Leftrightarrow&
                \true \OR I \leq_C J
\\
&\Leftrightarrow&
                \true
\end{eqnarray*}

\item[$I \not\in A$, $J \in A$]

\begin{eqnarray*}
I \not\in A &\Rightarrow& CA(I) = m + 1 + C(I)
\\
J \in A &\Rightarrow&  CA(J) = C(J)
\\
I \leq_{CA} J
\\
&\Leftrightarrow&
                CA(I) \leq CA(J)
\\
&\Leftrightarrow&
                m + 1 + C(I) \leq C(J)
\\
&\Leftrightarrow&
                \false
\\
\\
I \leq_A J
\\
&\Leftrightarrow&
                I \in A \OR J \not\in A
\\
&\Leftrightarrow&
                \false \OR \false
\\
&\Leftrightarrow&
                \false
\\
J \leq_A I
\\
&\Leftrightarrow&
                J \in A \OR I \not\in A
\\
&\Leftrightarrow&
                \true \OR \true
\\
&\Leftrightarrow&
                \true
\\
I \leq_A J \AND (J \not\leq_A I \OR I \leq_C J)
\\
&\Leftrightarrow&
                I \leq_A J \AND (\NOT J \leq_A I \OR I \leq_C J)
\\
&\Leftrightarrow&
                \false \AND (\NOT \true \OR I \leq_C J)
\\
&\Leftrightarrow&
                \false
\end{eqnarray*}

\item[$I \not\in A$, $J \not\in A$]

\begin{eqnarray*}
I \not\in A &\Rightarrow& CA(I) = m + 1 + C(I)
\\
J \not\in A &\Rightarrow& CA(I) = m + 1 + C(I)
\\
I \leq_{CA} J
\\
&\Leftrightarrow&
                CA(I) \leq CA(J)
\\
&\Leftrightarrow&
                m + 1 + C(I) \leq m + 1 + C(J)
\\
&\Leftrightarrow&
                C(I) \leq C(J)
\\
&\Leftrightarrow&
                I \leq_C J
\\
\\
I \leq_A J
\\
&\Leftrightarrow&
                I \in A \OR J \not\in A
\\
&\Leftrightarrow&
                \false \OR \true
\\
&\Leftrightarrow&
                \true
\\
J \leq_A I
\\
&\Leftrightarrow&
                J \in A \OR I \not\in A
\\
&\Leftrightarrow&
                \false \OR \true
\\
&\Leftrightarrow&
                \true
\\
I \leq_A J \AND (J \not\leq_A I \OR I \leq_C J)
\\
&\Leftrightarrow&
                I \leq_A J \AND (\NOT J \leq_A I \OR I \leq_C J)
\\
&\Leftrightarrow&
                \true \AND (\NOT \true \OR I \leq_C J)
\\
&\Leftrightarrow&
                \NOT \true \OR I \leq_C J
\\
&\Leftrightarrow&
                \false \OR I \leq_C J
\\
&\Leftrightarrow&
                I \leq_C J
\end{eqnarray*}

\end{description}

\long\def\ttytex#1#2{#1}
\ttytex{
}{
notation:
C(I) is the class index of I, the number i such that I c C(i)

        I <=C I    iff   C(I) <= C(J)

claim:
I <=Clex(A) J    iff    I <=A J AND (J /<=A I OR I <=C J)
        where Clex(A) = [C0A .. CmA C0-A .. Cm-A]

        I c A   =>  Clex(A)(I) = C(I)
        I /c A  =>  Clex(A)(I) = m + 1 + C(I)

four cases depending on IcA and JcA
the two sides of the claim are evaluated in each:
        1. I <=Clex(A) J
        2. I <=A J AND (J /<=A I OR I <=C J)

case
I c A, J c A

        I c A => Clex(A)(I) = C(I)
        J c A => Clex(A)(J) = C(J)
        I <=Clex(A) J =
                Clex(A)(I) <= Clex(A)(J)               : definition
                C(I) <= C(J)                           : I c A, J c A
                I <=C J                                : definition

        I <=A J
                I c A OR J /c A                        : definition
                true OR false
                true
        J <=A I
                J c A OR I /c A                        : definition
                true OR false
                true
        I <=A J AND (J /<=A I OR I <=C J)
                I <=A J AND (NOT J <=A I OR I <=C J)
                true AND (NOT true OR I <=C J)
                true AND (false OR I <=C J)
                false OR I <=C J
                I <=C J

case
I c A, J /c A

        I c A =>  Clex(A)(I) = C(I)
        J /c A => Clex(A)(J) = m + 1 + C(J)
        I <=Clex(A) J =
                Clex(A)(I) <= Clex(A)(J)               : definition
                C(I) <= m + 1 + C(J)                   : I c A, J /c A
                true                                   : C(I) <= m

        I <=A J
                I c A OR J /c A                        : definition
                true OR true
                true
        J <=A I
                J c A OR I /c A                        : definition
                false OR false
                false
        I <=A J AND (J /<=A I OR I <=C J)
                I <=A J AND (NOT J <=A I OR I <=C J)
                true AND (NOT false OR I <=C J)
                NOT false OR I <=C J
                true OR I <=C J
                true

case
I /c A, J c A
        I /c A => Clex(A)(I) = m + 1 + C(I)
        J c A =>  Clex(A)(J) = C(J)
        I <=Clex(A) J =
                Clex(A)(I) <= Clex(A)(J)               : definition
                m + 1 + C(I) <= C(J)                   : I /c A, J c A
                false                                  : C(J) <= m

        I <=A J
                I c A OR J /c A                        : definition
                false OR false
                false
        J <=A I
                J c A OR I /c A                        : definition
                true OR true
                true
        I <=A J AND (J /<=A I OR I <=C J)
                I <=A J AND (NOT J <=A I OR I <=C J)
                false AND (NOT true OR I <=C J)
                false

case
I /c A, J /c A
        I /c A => Clex(A)(I) = m + 1 + C(I)
        J /c A => Clex(A)(I) = m + 1 + C(I)
        I <=Clex(A) J =
                Clex(A)(I) <= Clex(A)(J)               : definition
                m + 1 + C(I) <= m + 1 + C(J)           : I /c A, J /c A
                C(I) <= C(J)
                I <=C J

        I <=A J
                I c A OR J /c A                        : definition
                false OR true
                true
        J <=A I
                J c A OR I /c A                        : definition
                       false OR true
                true
        I <=A J AND (J /<=A I OR I <=C J)
                I <=A J AND (NOT J <=A I OR I <=C J)
                true AND (NOT true OR I <=C J)
                NOT true OR I <=C J
                false OR I <=C J
                I <=C J
}

\qed

}

This equivalent formulation of the lexicographic revision of an order includes
the case where the order is generated by a sequence of other lexicographic
revisions.

\begin{corollary}
\label{lexicographic-sequence}

The doxastic state produced by the lexicographic revisions
{} $S = [S_1,\ldots,S_m]$
on the flat order $\emptyset$ compares $I \leq_{\emptyset S} J$ if:

\begin{itemize}

\item either $S=[]$ or

\item
\begin{itemize}

\item[] $I \leq_{S_m} J$ and

\item[]
either $J \not\leq_{S_m} I$
	or
$I \leq_{\emptyset [S_1,\ldots,S_{m-1}]} J$.

\end{itemize}

\end{itemize}

\end{corollary}

A property of the concatenation of sequences of lexicographic revisions is
proved.

\begin{definition}
\label{concat}

If $R=[R_1,\ldots,R_m]$ and $Q=[Q_1,\ldots,Q_m]$, then $R \cdot Q =
[R_1,\ldots,R_m,Q_1,\ldots,Q_m]$ is their concatenation.

\end{definition}

\state{break}{lemma}{}{

\begin{lemma}
\statelabel

If $S = R \cdot Q$, then $I \leq_{\emptyset S} J$ holds if and only if:

\begin{itemize}

\item $I \leq_{\emptyset Q} J$, and

\item either $J \not\leq_{\emptyset Q} I$ or $I \leq_{\emptyset R} J$.

\end{itemize}

\end{lemma}

}{

\proof The proof is by induction on the length of the sequence $Q$ only. The
length of $R$ does not matter.

\

The base case is $Q=[]$.

By definition, $I \leq_{\emptyset Q} J$ is always the case if $Q=[]$. As a
result, $J \not\leq_{\emptyset Q} I$ is never. The condition of the lemma
simplifies from
{} ``$I \leq_{\emptyset Q} J$ and
{}   either $J \not\leq_{\emptyset Q} I$ or $I \leq_{\emptyset R} J$''
to
{} ``$\true$ and either $\false$ or $I \leq_{\emptyset R} J$'',
which is the same as $I \leq_{\emptyset R} J$. This is also the same as $I
\leq_{\emptyset R \cdot Q} J$ since $Q=[]$.

\

In the induction case, $Q$ is not empty: it contains at least a formula $Q_m$,
possibly preceded by others $Q_1,\ldots,Q_{m-1}$. The induction assumption is
that the claim holds for every sequence shorter than $Q$, such as
$Q_1,\ldots,Q_{m-1}$.

The claim is that the condition in the statement of the lemma is equivalent to
$I \leq_{\emptyset R \cdot Q} J$. This condition is:

\begin{itemize}

\item $I \leq_{\emptyset Q} J$; and

\item either

\begin{itemize}

\item $J \not\leq_{\emptyset Q} I$ or

\item $I \leq_{\emptyset R} J$.

\end{itemize}

\end{itemize}

The first part $I \leq_{\emptyset Q} J$ is defined as
{} ``$Q=[]$ or
{}   $(I \leq_{\emptyset {Q_m}} J$ and
{}    ($J \not\leq_{\emptyset {Q_m}} I$ or $I \leq_{\emptyset Q'} J$)'', where
{} $Q'=[Q_1,\ldots,Q_{m-1}]$.
The first part $Q=[]$ is false because $Q$ is not empty.

\begin{itemize}

\item $I \leq_{\emptyset {Q_m}} J$; and

\item $J \not\leq_{\emptyset {Q_m}} I$ or $I \leq_{\emptyset R'} J$, and

\item either

\begin{itemize}

\item $J \not\leq_{\emptyset Q} I$ or

\item $I \leq_{\emptyset R} J$.

\end{itemize}

\end{itemize}

By definition, $J \not\leq_{\emptyset Q} I$ is the contrary of $J
\leq_{\emptyset Q} I$.

\begin{itemize}

\item $I \leq_{\emptyset {Q_m}} J$; and

\item $J \not\leq_{\emptyset {Q_m}} I$ or $I \leq_{\emptyset Q'} J$, and

\item either

\begin{itemize}

\item not $J \leq_{\emptyset Q} I$ or

\item $I \leq_{\emptyset R} J$.

\end{itemize}

\end{itemize}

Replacing $J \leq_{\emptyset Q} I$ with its definition leads to:

\begin{itemize}

\item $I \leq_{\emptyset {Q_m}} J$; and

\item $J \not\leq_{\emptyset {Q_m}} I$ or $I \leq_{\emptyset Q'} J$, and

\item either

\begin{itemize}

\item not(
$J \leq_{\emptyset {Q_m}} I$ and
either $I \not\leq_{\emptyset {Q_m}} J$ or $J \leq_{\emptyset Q'} I$
) or

\item $I \leq_{\emptyset R} J$.

\end{itemize}

\end{itemize}

The negation of a condition is the same as the negation of the elementary
subconditions it is made of when swapping conjunctions and disjunctions.

\begin{itemize}

\item $I \leq_{\emptyset {Q_m}} J$; and

\item $J \not\leq_{\emptyset {Q_m}} I$ or $I \leq_{\emptyset Q'} J$, and

\item either

\begin{itemize}

\item $J \not\leq_{\emptyset {Q_m}} I$ or
       both $I \leq_{\emptyset {Q_m}} J$ and $J \not\leq_{\emptyset Q'} I$, or

\item $I \leq_{\emptyset R} J$.

\end{itemize}

\end{itemize}

The latter two subconditions form a single disjunction:

\begin{itemize}

\item $I \leq_{\emptyset {Q_1}} J$; and

\item either

\begin{itemize}

\item $J \not\leq_{\emptyset {Q_1}} I$ or

\item $I \leq_{\emptyset Q'} J$ and

\end{itemize}

\item either

\begin{itemize}

\item $J \not\leq_{\emptyset {Q_m}} I$ or

\item both $I \leq_{\emptyset {Q_m}} J$ and $J \not\leq_{\emptyset Q'} I$, or

\item $I \leq_{\emptyset R} J$.

\end{itemize}

\end{itemize}

The first line states the truth of $I \leq_{\emptyset {Q_m}} J$. Therefore, it
can be removed from the other line that contains it.

\begin{itemize}

\item $I \leq_{\emptyset {Q_m}} J$; and

\item either

\begin{itemize}

\item $J \not\leq_{\emptyset {Q_m}} I$ or

\item $I \leq_{\emptyset Q'} J$ and

\end{itemize}

\item either

\begin{itemize}

\item $J \not\leq_{\emptyset {Q_m}} I$ or

\item $J \not\leq_{\emptyset Q'} I$, or

\item $I \leq_{\emptyset R} J$.

\end{itemize}

\end{itemize}

The comparison $J \not\leq_{\emptyset {Q_m}} I$ is in both the second and the
third point. It can be factored out.


\begin{itemize}

\item $I \leq_{\emptyset {Q_m}} J$; and

\item either

\begin{itemize}

\item $J \not\leq_{\emptyset {Q_m}} I$, or

\item both

\begin{itemize}

\item $I \leq_{\emptyset Q'} J$ and

\item $J \not\leq_{\emptyset Q'} I$ or $I \leq_{\emptyset R} J$.

\end{itemize}

\end{itemize}

\end{itemize}

Since $Q'$ is strictly shorter than $Q$, the induction assumption applies: $I
\leq_{\emptyset R \cdot Q'} J$ is equivalent to the condition in the statement
of the lemma when applied to $Q'$ instead of $R$:
{} ``$I \leq_{\emptyset Q'} J$ and
{}   either $J \not\leq_{\emptyset Q'} I$ or $I \leq_{\emptyset R} J$''.
This is the same as the latter two points above.

\begin{itemize}

\item $I \leq_{\emptyset {Q_m}} J$; and

\item either

\begin{itemize}

\item $J \not\leq_{\emptyset {Q_m}} I$, or

\item $I \leq_{\emptyset R \cdot Q'} J$

\end{itemize}

\end{itemize}

This condition defines $I \leq_{\emptyset R \cdot Q} J$.~\qed

}

A simple property follows: since $I \leq_{\emptyset R \cdot Q} J$ is equivalent
to $I \leq_{\emptyset Q} J$ and some other condition, the former implies the
latter.

\begin{corollary}
\label{monotone}

$I \leq_{\emptyset R \cdot Q} J$ implies $I \leq_{\emptyset Q} J$

\end{corollary}

For connected preorders, the strict order $I < J$ is the same as the negation
of the order on swapped elements $J \not\leq I$. The corollary is therefore the
same as $I <_{\emptyset Q} J$ implying $I <_{\emptyset R \cdot Q} J$: a strict
order induced by any sequence of revisions is never reversed by prefixing the
sequence by earlier revisions. If $I <_{\emptyset Q} J$ is the case, then $I
<_{\emptyset R \cdot Q} J$ is also the case, no matter what the previous
formulae $R$ say. This is in particular the case when $R$ consists the single
formula to be checked for redundancy.

Lemma~\ref{break} and Corollary~\ref{monotone} simplify the condition of
equality of orders of a sequence and its suffixes.

\state{equivalent-made-not}{lemma}{}{

\begin{lemma}
\statelabel

The order $\emptyset S \cdot R$ is not the same as $\emptyset R$ if and only if
there exists models $I$ and $J$ such that $I \leq_{\emptyset R} J$, $J
\leq_{\emptyset R} I$ and $I \not\leq_{\emptyset S} J$ all hold. Equivalently,
there exist two models $I, J$ equivalent according to $S$ such that
$I\not\leq_R J$.

\end{lemma}

}{

\proof By definition, two orders differ if and only if they compare the same
two models differently. The two cases are:

\begin{itemize}

\item $I \leq_{\emptyset S \cdot R} J$ and $I \not\leq_{\emptyset R} J$, or

\item $I \not\leq_{\emptyset S \cdot R} J$ and $I \leq_{\emptyset R} J$.

\end{itemize}

Corollary~\ref{monotone} excludes the first case, since 
{} $I \leq_{\emptyset S \cdot R} J$ implies $I \leq_{\emptyset R} J$.
Only the second may hold:

\[
I \not\leq_{\emptyset S \cdot R} J \mbox{ and } I \leq_{\emptyset R} J
\]

This is the same as the following two conditions.

\begin{itemize}

\item $I \leq_{\emptyset R} J$ and

\item not $(I \leq_{\emptyset S \cdot R} J)$

\end{itemize}

Lemma~\ref{break} rewrites the second.

\begin{itemize}

\item $I \leq_{\emptyset R} J$ and

\item not ($I \leq_{\emptyset R} J$ and
      either $J \not\leq_{\emptyset R} I$ or $I \leq_{\emptyset S} J$)

\end{itemize}

The negation can be pushed in by swapping conjunctions and disjunctions.

\begin{itemize}

\item $I \leq_{\emptyset R} J$ and

\item $I \not\leq_{\emptyset R} J$ or
      both $J \leq_{\emptyset R} I$ and $I \not\leq_{\emptyset S} J$

\end{itemize}

The first line states $I \leq_{\emptyset R} J$. As a result, $I
\not\leq_{\emptyset R} J$ is false. Since it is a case of a disjunction, it can
be removed.

\begin{itemize}

\item $I \leq_{\emptyset R} J$ and

\item $J \leq_{\emptyset R} I$ and $I \not\leq_{\emptyset S} J$

\end{itemize}

This is a conjunction of three conditions.

\[
I \leq_{\emptyset R} J
\mbox{ and }
J \leq_{\emptyset R} I
\mbox{ and }
I \not\leq_{\emptyset S} J
\]

A chain of equivalences led from the difference of $\emptyset S \cdot R$ and
$\emptyset R$ to the existence of two models $I$ and $J$ meeting this
condition.~\qed

}

A simple consequence on redundancy follows.

\begin{corollary}
\label{equivalent-made-not-single}

The order $\emptyset [F] \cdot S$ differs from $\emptyset S$ if and only if $I
\leq_{\emptyset S} J$, $J \leq_{\emptyset S} I$ and $I \not\leq_{\emptyset F}
J$ hold for some pair of models $I$ and $J$.

\end{corollary}

Equivalence between models has a simple formulation in terms of the splittings
of the sequence.

\state{product}{lemma}{}{

\begin{lemma}
\statelabel

The comparison $I \equiv_{\emptyset S \cdot R} J$
is equivalent to the truth of both 
$I \equiv_{\emptyset S} J$ and $I \equiv_{\emptyset R} J$.

\end{lemma}

}{

\proof The equivalence $I \equiv_{\emptyset S \cdot R} J$ is defined as the
truth of the comparison in both directions.

\begin{itemize}

\item $I \leq_{\emptyset S \cdot R} J$ and

\item $J \leq_{\emptyset S \cdot R} I$

\end{itemize}

Lemma~\ref{break} rewrites both comparisons.

\begin{itemize}

\item $I \leq_{\emptyset R} J$ and
      either $J \not\leq_{\emptyset R} I$ or $I \leq_{\emptyset S} J$ and

\item $J \leq_{\emptyset R} I$ and
      either $I \not\leq_{\emptyset R} J$ or $J \leq_{\emptyset S} I$

\end{itemize}

The conjunction of two conjunctions is a single conjunction.

\begin{itemize}

\item $I \leq_{\emptyset R} J$ and

\item either $J \not\leq_{\emptyset R} I$ or $I \leq_{\emptyset S} J$ and

\item $J \leq_{\emptyset R} I$ and

\item either $I \not\leq_{\emptyset R} J$ or $J \leq_{\emptyset S} I$

\end{itemize}

The condition $I \leq_{\emptyset R} J$ in the first line negates $I
\not\leq_{\emptyset R} J$ in the fourh. The condition $J \leq_{\emptyset R} I$
of the third line negates $J \not\leq_{\emptyset R} I$ in the second. The
second and the fourth lines are disjunctions. A false conjunct is removable.

\begin{itemize}

\item $I \leq_{\emptyset R} J$ and

\item $I \leq_{\emptyset S} J$ and

\item $J \leq_{\emptyset R} I$ and

\item $J \leq_{\emptyset S} I$

\end{itemize}

The first and third line define $I \equiv_{\emptyset R} J$. The second and the
fourth define $I \equiv_{\emptyset S} J$.

\begin{itemize}

\item $I \equiv_{\emptyset R} J$ and

\item $I \equiv_{\emptyset S} J$

\end{itemize}

This is the claim of the lemma.~\qed

}

Equivalence of a concatenation is the same as equivalence of the two parts. The
specialization to redundancy is that $I \equiv_{\emptyset [F] \cdot S} J$ is
the same as $I \equiv_{\emptyset S} J$ and $I \equiv_{\emptyset F} J$.

The check of equivalence $I \equiv_{\emptyset S} J$ can be split among
conjunctions of the formulae of the sequence $S$, some negated. Namely, $I
\equiv_{\emptyset S} J$ is the same as $I \equiv_{\emptyset Q} J$ for all
conjunctions $Q$ of all formulae of $S$, each possibly negated. For example, $I
\equiv_{\emptyset [S_1,S_2,S_3]} J$ is the same as $I \equiv_{\emptyset Q} J$
for all formulae $Q$ like $S_1 \wedge \neg S_2 \wedge S_3$, with negations
prepended to the formulae in all possible ways.

Formally, such conjunctions are written
{} $Q = (B_1 \equiv S_1) \wedge \cdots \wedge (B_m \equiv S_m)$
where each $B_i$ is either $\true$ or $\false$. This is mathematical expression
of a conjunction of all formulae of $S_1,\ldots,S_m$ where each formula can be
prepended with a negation $\neg$ or not. For example,
{} $Q = (\true \equiv S_1) \wedge (\false \equiv S_2) \wedge (\true \equiv S_3)$
is the same as
{} $Q = S_1 \wedge \neg S_2 \wedge S_3$
since $\true \equiv S_1$ is equivalent to $S_1$ and $\false \equiv S_2$ to
$\neg S_2$ and $\true \equiv S_3$ to $S_3$.

\begin{definition}
\label{q-combination}

A Q-combination of formulae $S_1,\ldots,S_m$ is a formula
{} $Q = (B_1 \equiv S_1) \wedge \cdots \wedge (B_m \equiv S_m)$
where each $B_i$ is either $\true$ or $\false$.

\end{definition}

Equivalence on all Q-combinations is the same as equivalence on the sequence.
This is proved by a sequence of lemmas. The first is that $I \equiv_{\emptyset
Q} J$ for all Q-combinations $Q$ implies $I \equiv_{\emptyset S_i} J$ for all
formulae $S_i$.

\state{conjunction-single}{lemma}{}{

\begin{lemma}
\statelabel

If $I \equiv_{\emptyset Q} J$ holds for every formula
{} $Q = (B_1 \equiv S_1) \wedge \cdots \wedge (B_m \equiv S_m)$
such that every $B_i$ is either $\true$ or $\false$,
then $I \equiv_{\emptyset S_i} J$ holds for every $i \in \{1,\ldots,m\}$.

\end{lemma}

}{

\proof Since $I \equiv_{\emptyset Q} J$ holds for every formula
{} $Q = (B_1 \equiv S_1) \wedge \cdots \wedge (B_m \equiv S_m)$
such that every $B_i$ is either $\true$ or $\false$, it also holds for the
following values of $B_i$:

\begin{description}

\item[$B_i = \true$] if $I \models S_i$;

\item[$B_i = \false$] if $I \not\models S_i$.

\end{description}

The resulting formula $Q$ is shown to be satisfied by $I$.

If $I \models S_i$ then $B_i = \true$. Therefore, $B_i \equiv S_i$ is
equivalent to $\true \equiv S_i$, which is equivalent to $S_i$. As a result, $I
\models B_i \equiv S_i$.

If $I \not\models S_i$ then $B_i = \false$. Therefore, $B_i \equiv S_i$ is
equivalent to $\false \equiv S_i$, which is equivalent to $\neg S_i$. As a
result, $I \models B_i \equiv S_i$.

Since $I$ satisfies all formulae $B_i \equiv S_i$ regardless of $I \models S_i$
or $I \not\models S_i$, it satisfies all of them. As a result, it satisfies
$Q$.

The assumption includes $I \equiv_{\emptyset Q} J$ for this specific formula
$Q$. This condition includes $J \leq_{\emptyset Q} I$, which is defined as
{} $J \models Q$ or $I \not\models Q$.
Since $I$ is proved to satisfy $Q$, only the first case $J \models Q$ is
possible.

Since $J$ satisfies the conjunction $Q$, it satisfies all its components $B_i
\equiv S_i$. Replacing $B_i$ by its value in $J \models B_i \equiv S_i$ gives:

\begin{description}

\item[$J \models \true \equiv S_i$] if $I \models S_i$;

\item[$J \models \false \equiv S_i$] if $I \not\models S_i$.

\end{description}

In other words, if $I \models S_i$ then $J \models S_i$, and if $I \not\models
S_i$ then $J \not\models S_i$.~\qed

}

The equivalence of $I \equiv_{\emptyset S} J$ and $I \equiv_{\emptyset Q} J$
for all formulae $Q$ can now be proved in one direction: from $Q$ to $S$.

\state{conjunctions-equivalent}{lemma}{}{

\begin{lemma}
\statelabel

If $S=[S_1,\ldots,S_m]$ and $I \equiv_{\emptyset Q} J$ holds for all formulae
{} $Q = (B_1 \equiv S_1) \wedge \cdots \wedge (B_m \equiv S_m)$
where every $B_i$ is either $\true$ or $\false$, then $I \equiv_{\emptyset S}
J$.

\end{lemma}

}{

\proof Lemma~\ref{conjunction-single} proves $I \equiv_{\emptyset S_i} J$ for
every formula $S_i$ under the assumption of this lemma. This equivalence $I
\equiv_{\emptyset S_i} J$ is defined as $I \leq_{\emptyset S_i} J$ and $J
\leq_{\emptyset S_i} I$, and holds for every formula $S_i$ of $S$.

The claim $I \leq_{\emptyset S} J$ follows from $I \leq_{\emptyset S_i} J$ for
every $S_i$ by induction on the length of $S$. In the base case, $S$ is empty
and $I \leq_{\emptyset S} J$ holds by definition. If $S$ is not empty, $I
\leq_{\emptyset R} J$ holds where $R$ is the sequence of the formulae of $S$
but its last one by the induction assumption. The conclusion to prove $I
\leq_{\emptyset S} J$ is defined as
{} $I \leq_{\emptyset S_m} J$ and
{}  either $J \not\leq_{\emptyset S_m} I$ or $I \leq_{\emptyset R} J$.
The first subcondition $I \leq_{\emptyset S_m} J$ holds because $I
\leq_{\emptyset S_i} J$ holds for every formula $S_i$ of $S$, including $S_m$.
The second subcondition
{} ``either $J \not\leq_{\emptyset S_m} I$ or $I \leq_{\emptyset R} J$''
holds because its second part $I \leq_{\emptyset R} J$ holds. This proves $I
\leq_{\emptyset S} J$ in the induction case.

The conclusion is that $I \leq_{\emptyset S_i} J$ for all formulae $S_i$ of $S$
implies $I \leq_{\emptyset S} J$. For the same reason, $J \leq_{\emptyset S_i}
I$ for all formulae $S_i$ of $S$ implies $J \leq_{\emptyset S} I$.

The required conclusion $I \equiv_{\emptyset S} J$ is therefore proved.~\qed

}

The other direction is shown by the following lemma: from $S$ to all formulae
$Q$.

\state{equivalent-conjunctions}{lemma}{}{

\begin{lemma}
\statelabel

If $I \equiv_{\emptyset S} J$ holds, then $I \equiv_{\emptyset Q} J$ holds for
every formula $Q = (B_1 \equiv S_1) \wedge \cdots \wedge (B_m \equiv S_m)$,
where every $B_i$ is either $\true$ or $\false$ and $S=[S_1,\ldots,S_m]$.

\end{lemma}

}{

\proof Proof is by induction on the length of $S$.

The base case is $S=[]$. Since $S$ does not contain any formula, the only
formula $Q$ is the empty conjunction: $Q=\true$. In such conditions, $I
\leq_{\emptyset S} J$ holds by definition. The equivalence $I \equiv_{\emptyset
Q} J$ holds because $I \models Q$ implies $I \leq_{\emptyset Q} J$ and $J
\models Q$ implies $J \leq_{\emptyset Q} I$.

In the induction case, $S$ is a non-empty sequence $[S_1,\ldots,S_m]$ where the
first element $S_1$ exists. By construction, the claim holds for every sequence
shorter than $S$, such as
{} $R = [S_2,\ldots,S_m]$.
By construction, $S = [S_1] \cdot R$

The premise of the lemma is $I \equiv_{\emptyset S} J$. Lemma~\ref{product}
states that $I \equiv_{\emptyset S} J$ is equivalent $I \equiv_{\emptyset
[S_1]} J$ and $I \equiv_{\emptyset R} J$ and therefore implies both.

Let $B_1,\ldots,B_m$ be an arbitrary $m$-tuple of values in $\{\true,\false\}$.

By the inductive assumption, the claim holds for both $[S_1]$ and $R$.
Therefore, $I \equiv_{\emptyset Q} J$ holds where $Q = B_1 \equiv S_1$ and $I
\equiv_{\emptyset Q'} J$ holds where $Q' = (B_2 \equiv S_2) \wedge \cdots
\wedge (B_m \equiv S_m)$.

Such equivalences mean that $Q$ evaluates the same in $I$ and $J$, and the same
does $Q'$. Since $Q$ and $Q'$ both evaluate the same in $I$ and $J$, also $Q
\wedge Q'$ does. This conjunction is
{} $Q = (B_1 \equiv S_1) \wedge
{}      (B_2 \equiv S_2) \cdots \wedge (B_m \equiv S_m)$,
which evaluates the same in $I$. Technically, $I \equiv_{\emptyset Q} J$.

This equivalence $I \equiv_{\emptyset Q} J$ is the case for every $m$-tuples of
values $B_1,\ldots,B_m$, which is the claim.~\qed

}

Since both directions are proved, the condition is necessary and sufficient.

\state{conjunctions}{theorem}{}{

\begin{theorem}
\statelabel

The equivalence $I \equiv_{\emptyset S} J$ holds iff and only if $I
\equiv_{\emptyset Q} J$ holds for every formula $Q = (B_1 \equiv S_1) \wedge
\cdots \wedge (B_m \equiv S_m)$, where every $B_i$ is either $\true$ or
$\false$ and $S = [S_1,\ldots,S_m]$.

\end{theorem}

}{

\proof The two directions of the claim are proved by
Lemma~\ref{equivalent-conjunctions} and
Lemma~\ref{conjunctions-equivalent}.~\qed

}

\subsection{Redundancy of a sequence of lexicographic revisions}

The theorem ending the previous section reformulates the equivalence of two
models according to a sequence: $I \equiv_{\emptyset S} J$. The final
destination is a reformulation of the equality of $\emptyset
[S_1,S_2,\ldots,S_m]$ and $\emptyset [S_2,\ldots,S_m]$, the redundancy of the
first formula of a sequence from the flat doxastic state.

The first lemma in this path is about the formulae $Q$. By construction, if
they differ even by a single $B_i$, then one contains a formula $S_i$ and the
other $\neg S_i$; therefore, they do not share models.

\state{no-share}{lemma}{}{

\begin{lemma}
\statelabel

Two Q-combinations $Q$ and $Q'$ of $S_1,\ldots,S_m$ do not share models if
there exists $i$ such that $B_i \not= B_i'$, where:
\begin{eqnarray*}
Q &=& (B_1 \equiv S_1) \wedge \cdots \wedge (B_{m-1} \equiv S_{m-1})	\\
Q' &=& (B_1' \equiv S_1) \wedge \cdots \wedge (B_{m-1}' \equiv S_{m-1})
\end{eqnarray*}

\end{lemma}

}{



\proof The latter premise is that at least a $B_i$ differs from $B_i'$. Since
$B_i$ and $B_i'$ can only be $\true$ or $\false$, if one is $\true$ the other
is $\false$ and vice versa.

In the first case, $B_i \equiv S_i$ is $\true \equiv S_i$, which is equivalent
to $S_i$, and $B_i' \equiv S_i$ is $\false \equiv S_i$, which is equivalent to
$\neg S_i$.

In the second case, $B_i \equiv S_i$ is $\false \equiv S_i$, which is
equivalent to $\neg S_i$, and $B_i' \equiv S_i$ is $\true \equiv S_i$, which is
equivalent to $S_i$.

In both cases, $B_i' \equiv S_i$ is equivalent to the negation of $B_i \equiv
S_i$. If a model satisfies $Q$ then it satisfies its conjunct $B_i \equiv S_i$
and therefore falsifies $B_i' \equiv S_i$. Since $B_i' \equiv S_i$ is a
conjunct of $Q'$, this formula is also falsified by $I$.~\qed

}

Redundancy is expressed in terms of the Q-combinations by the following
theorem.

\state{relevance}{theorem}{}{

\begin{theorem}
\statelabel

The order $\emptyset [S_1,S_2,\ldots,S_m]$ coincides with $\emptyset
[S_2,\ldots,S_m]$ if and only if either $Q \models S_1$ or $Q \models \neg S_1$
hold for every Q-combination of $S_2,\ldots,S_m$.

\end{theorem}

}{

\proof Definining $S' = [S_2,\ldots,S_m]$ and $S = [S_1] \cdot S'$, the
claim is a consequence of Corollary~\ref{equivalent-made-not-single} and
Theorem~\ref{conjunctions} when applied to $S$, $S'$ and $S_1$:
\begin{eqnarray*}
\emptyset [S_1] \cdot S' \not= \emptyset S'
&\mbox{ is equivalent to }&
\exists I,J ~.~ I \equiv_{\emptyset S'} J, I \not\leq_{\emptyset S_1} J
\\
I \equiv_{\emptyset S'} I
&\mbox{ is equivalent to }&
\forall B_2,\ldots,B_m \in \{\true,\false\} ~.~
I \equiv_{\emptyset Q} J,
\\
&& Q = (B_2 \equiv S_2) \wedge \cdots \wedge (B_m \equiv S_m)
\end{eqnarray*}

The claim is proved in reverse: $\emptyset S = \emptyset S'$ is false if and
only if
{} $Q = (B_2 \equiv S_2) \wedge \cdots \wedge (B_m \equiv S_m)$
entails neither $S_1$ nor $\neg S_1$ for some values of $B_2,\ldots,B_m$ in
$\{\false,\true\}$.


Corollary~\ref{equivalent-made-not-single} reformulates the falsity of
$\emptyset S = \emptyset S'$ as the existence of two models $I$ and $J$ such
that $I \equiv_{\emptyset S'} J$ and $I \not\leq_{\emptyset S_1} J$. This is
proved to be the same to the existence of a formula
{} $Q = (B_2 \equiv S_2) \wedge \cdots \wedge (B_m \equiv S_m)$
such that $Q \not\models S_1$ and $Q \not\models \neg S_1$ where each $B_i$ is
either $\true$ or $\false$.


%

%


\

The first direction of the claim assumes the existence of a pair of models $I$
and $J$ such that $I \equiv_{\emptyset S'} J$ and $I \not\leq_{\emptyset S_1}
J$.

Theorem~\ref{conjunctions} proves that $I \equiv_{\emptyset S'} J$ entails $I
\equiv_{\emptyset Q} J$ for every formula
{} $Q = (B_1 \equiv S_1) \wedge \cdots \wedge (B_{m-1} \equiv S_{m-1})$
such that each $B_i$ is either $\true$ or $\false$.

In particular, $I \equiv_{\emptyset Q} J$ holds for the formula $Q$ stemming
from the specific values $B_i = \true$ if $I \models S_1$ and $B_i = \false$
otherwise.

Every conjunct $B_i \equiv S_i$ of $Q$ is satisfied by $I$: if $I \models S_i$
then $B_i = \true$, which makes $B_i \equiv S_i$ equivalent to $S_i$; if $I
\models\neg S_i$ then $B_i = \false$, which makes $B_i \equiv S_i$ equivalent
to $\neg S_i$. This is the case for every $S_i \in S'$, making $I \models Q$
the case.

Since $I \equiv_{\emptyset Q} J$ implies $J \models_{\emptyset Q} I$, which is
$J \models Q$ or $I \not\models Q$, it implies $J \models Q$ since $I
\not\models Q$ is false. This is a consequence of the first assumption $I
\equiv_{\emptyset S'} J$.

The second assumption $I \not\leq_{\emptyset S_1} J$ implies $I \not\models
S_1$ and $J \models S_m$. Overall, $I$ and $J$ satisfy $Q$ while $I$ satisfies
$\neg S_1$ and $J$ satisfies $S_1$. The model $I$ is a counterexample to $Q
\models S_1$, the model $J$ is a counterexample to $Q \models \neg S_1$.

The first direction of the claim is proved: $Q \not\models S_1$ and $Q
\not\models \neg S_1$ hold for some formula
{} $Q = (B_2 \equiv S_2) \wedge \cdots \wedge (B_m \equiv S_m)$
where each $B_i$ is either $\true$ or $\false$.

\

The other direction assumes that $Q \not\models S_1$ and $Q \not\models \neg
S_1$ hold for some formula
{} $Q = (B_2 \equiv S_2) \wedge \cdots \wedge (B_m \equiv S_m)$
where each $B_i$ is either $\true$ or $\false$.

Since $Q \not\models S_1$, a model $I$ satisfies $Q$ but not $S_1$. Since $Q
\not\models \neg S_1$, a model $J$ satisfies $Q$ but not $\neg S_1$, which
implies that $J$ satisfies $S_1$. The conditions $I \not\models S_1$ and $J
\models S_1$ negate $I \leq_{\emptyset S_1} J$. This is the second part of the
claim: $I \not\leq_{\emptyset S_1} J$.

The first part of the claim is $I \equiv_{\emptyset S'} J$.
Theorem~\ref{conjunctions} proves it the same as $I \equiv_{\emptyset Q'} J$
for all formulae
{} $Q' = (B_2' \equiv S_2) \wedge \cdots \wedge (B_m' \equiv S_m)$
such that each $B_i$ is either $\true$ or $\false$. This is proved to be a
consequence of $I \models Q$ and $J \models Q$ for the specific formula $Q$ of
the assumption. The models $I$ and $J$ satisfy $Q$.

If $Q'$ is equal to $Q$ then $I \models Q$ and $J \models Q$ respectively imply
$I \models Q'$ and $J \models Q'$, which imply $I \equiv_{\emptyset Q'} J$.

If $Q'$ differs from $Q$, then Lemma~\ref{no-share} proves that $I \models Q$
and $J \models Q'$ respectively entail $I \not\models Q'$ and $J \not\models
Q'$. A consequence is $I \equiv_{\emptyset Q'} J$.

This proves $I \equiv_{\emptyset Q'} J$ for all formulae $Q$ regardless of
whether $Q'$ is the same as $Q$ or not. This is equivalent to $I
\equiv_{\emptyset S'} J$ by Theorem~\ref{conjunctions}.

For the chosen models $I$ and $J$, both $I \equiv_{\emptyset S'} J$ and $I
\not\leq_{\emptyset S_1} J$ hold. This is the claim.~\qed

}

Another reformulation of redundancy is equivalence with a disjunction of
Q-combinations.

\state{relevance-equivalence}{theorem}{}{

\begin{theorem}
\statelabel

The orders $\emptyset [S_1,S_2,\ldots,S_m]$ and $\emptyset [S_2,\ldots,S_m]$
coincide if and only if $S_1$ is equivalent to the disjunction of some
Q-combinations of $S_2,\ldots,S_m$.

\end{theorem}

}{

\proof Theorem~\ref{relevance} proves that the equality of the two orders is
equivalent to $Q \models S_m$ or $Q \models \neg S_1$ for every formula
{} $Q = (B_2 \equiv S_2) \wedge \cdots \wedge (B_m \equiv S_m)$,
where each $B_i$ is either $\true$ or $\false$. This condition is proved to be
equivalent to the equivalence between $S_1$ and the disjunction $D$ of the
formulae $Q$ such that $Q \models S_1$.

\



The first direction of the claim is that $Q \models S_1$ or $Q \models \neg
S_1$ for all formulae $Q$ implies the equivalence betwen $D$ and $S_1$.

It entails $S_1$ because each of its disjuncts $Q$ entails $S_1$.

Remains to prove the converse: $S_1$ entails $D$. This condition $S_1 \models
D$ is the same as $\neg D \models \neg S_1$: each model that falsifies $D$ also
falsifies $S_1$. A model $I$ falsifies the disjunction $D$ if and only if it
falsifies all its disjuncts.

A particular choice of $B_i$ is: $B_i = \true$ if $I \models S_i$ and $B_i =
\false$ otherwise. If $I \models S_i$ then $B_i = \true$; therefore, $B_i
\equiv S_i$ is $\true \equiv S_i$, which is equivalent to $S_i$, which is
satisfied by $I$. If $I \models \neg S_i$ then $B_i = \false$; therefore, $B_i
\equiv S_i$ is $\false \equiv S_i$, which is equivalent to $\neg S_i$, which is
satisfied by $I$. Since $I$ satisfies all formulae $B_i \equiv S_i$, it also
satisfies their conjunction.

This conjunction $Q$ is satisfied by $I$. It is not in $D$ because all
disjuncts of $D$ are falsified by $I$. Since $D$ comprises all formulae that
entail $S_1$ and $Q$ is not in it, $Q$ does not entail $S_1$. By assumption, it
entails $\neg S_1$. Since $I$ satisfies $Q$, it satisfies $\neg S_1$.

Since $I$ is an arbitrary model of $\neg D$ and has been proved to entail $\neg
S_1$, the conclusion is that $\neg D \models \neg S_1$, which is the same as
$S_1 \models Q$.

Both parts of the first direction of the claim are proved: $D \models S_1$ and
$S_1 \models Q$.

\



The second direction of the claim is that the equivalence between $S_1$ and a
disjunction $D$ of some formulae $Q$ implies $Q \models S_1$ or $Q \models \neg
S_1$ for all formulae $Q$.

The assumption is $S_1 \equiv D$ where $D$ is a disjunction of some formulae
$Q$, each being
{} $Q = (B_2 \equiv S_2) \wedge \cdots \wedge (B_m \equiv S_m)$
where each $B_i$ is either $\true$ or $\false$.

Let $Q$ be an arbitrary such formula, in $D$ or not. The two cases are analyzed
separately: $Q$ in $D$; $Q$ not in $D$.

The first case is that $Q$ is in $D$. Since $D$ is a disjunction of formulae
$Q$, if $Q$ is in $D$ it entails $D$. Since $D$ is equivalent to $S_1$, it
entails it. The chain $Q \models D$ and $D \models S_1$ leads to $Q \models
S_1$.

The second case is that $Q$ is not in $D$. The claim is that every model $I$ of
$Q$ satisfies $\neg S_1$.

Since $Q$ is not in $D$, every element $Q'$ of $D$ differs from $Q$.
Lemma~\ref{no-share} proves that $Q$ and $Q'$ do not share models. Since $I$ is
a model of $Q$, it is not a model of $Q'$. This is the case for all formulae
$Q'$ of $D$. Since $D$ is a disjunction of the formulae $Q'$ it contains, it
has exactly the union of their models. Since $I$ is not a model of any $Q'$, it
is not a model of $D$.

It is a model of $\neg D$, then. Since $D$ is equivalent to $S_m$, its negation
$\neg D$ is equivalent to $\neg S_m$. The conclusion is that $I$ is a model of
$\neg S_1$. Being $I$ an arbitrary model of $Q$, this proves that every model
of $Q$ satisfies $\neg S_1$.

This concludes the second direction of the proof: either $Q \models S_1$ or $Q
\models \neg S_1$.~\qed

}

Sequences of two formulae will prove an interesting subcase in the
computational analysis. Redundancy specializes as follows.

\state{relevance-two}{theorem}{}{

\begin{theorem}
\statelabel

The equality of $\emptyset [S_1,S_2]$ and $\emptyset [S_2]$ holds if and only
if either one of the following four conditions hold:

\begin{enumerate}

\item $S_1$ is inconsistent;

\item $S_1$ is tautological;

\item $S_1$ is equivalent to $S_2$;

\item $S_1$ is equivalent to $\neg S_2$.

\end{enumerate}

\end{theorem}

}{

\proof Theorem~\ref{relevance-equivalence} expresses the equality of $\emptyset
[S_1,S_2]$ and $\emptyset [S_2]$ as the equivalence of $S_1$ with the
disjunction of some formulae $Q = (B_2 \equiv S_2)$ where $B_2$ is either
$\true$ or $\false$.

Only two such formulae $Q$ exist:
{} $Q_1 = (\true \equiv S_2)$, which is equivalent to $S_2$,
and
{} $Q_2 = (\false \equiv S_2)$, which is equivalent to $\neg S_2$.
The four possible subsets of $\{Q_1, Q_2\}$ are considered in turn:
$\emptyset$, $\{S_2\}$, $\{\neg S_2\}$, and $\{S_2, \neg S_2\}$.

\begin{description}

\item[$\emptyset$]:

the disjunction $\vee \emptyset$ is $\false$; the condition that this
disjunction is equivalent to $S_2$ is the same as the inconsistency of $S_2$;

\item[$\{S_2\}$]:

the disjunction $\vee \{S_2\}$ is $S_2$; the condition that this disjunction is
equivalent to $S_1$ is the same as $S_1 \equiv S_2$;

\item[$\{\neg S_2\}$]:

the disjunction $\vee \{\neg S_2\}$ is $\neg S_2$; the condition that this
disjunction is equivalent to $S_1$ is the same as $S_1 \equiv \neg S_2$;

\item[$\{S_2,\neg S_2\}$]:

the disjunction $\vee \{S_2, \neg S_2\}$ is $S_2 \vee \neg S_2$, which is a
tautology; this disjunction is equivalent to $S_1$ if and only if $S_1$ is
tautological.

\end{description}

This proves that $\emptyset [S_1,S_2] = \emptyset [S_2]$ is the same as either
of the four conditions in the statement of the lemma: $S_1$ is either
inconsistent, tautological, equivalent to $S_2$ or equivalent to $\neg
S_2$.~\qed

}

This theorem ends the quest for a necessary and sufficient condition to
redundancy.

\subsection{Necessary conditions for redundancy}

Two other conditions are provided. They are necessary for redundancy, but not
sufficient.

The first is a form of monotony: if the first formula of a sequence is
redundant, it remains so even if other formulae are inserted after it. In terms
of information loss, a redundant formula can be removed without any fear that
it will turn important by a following revision.

\state{last-preceding}{theorem}{}{

\begin{theorem}
\statelabel

If
{} $\emptyset [F,S_1,\ldots,S_m] = \emptyset [S_1,\ldots,S_m]$,
then
{} $\emptyset [F,S_0,S_1,\ldots,S_m] = [\emptyset S_0,S_1,\ldots,S_m]$.

\end{theorem}

}{

\proof The premise of the theorem is recast according to
Theorem~\ref{relevance}: $Q \models F$ or $Q \models \neg F$ for every $Q$ as
follows, where each $B_i$ is either $\true$ or $\false$. The conclusion is the
same for $Q'$.
\begin{eqnarray*}
Q &=&	(B_1 \equiv S_1) \wedge \cdots \wedge (B_m \equiv S_m)	\\
Q' &=&	(B_0 \equiv S_0) \wedge
	(B_1 \equiv S_1) \wedge \cdots \wedge (B_m \equiv S_m) 
\end{eqnarray*}

Each formula $Q'$ is the same as
{} $(B_0 \equiv S_0) \wedge Q$
for some formula $Q$. The premise of the theorem is that $Q$ entails $F$ or
$\neg F$. Either way, since $Q'$ is a conjunction containing $Q$, it entails
all the formulae $Q$ entails, including either $F$ or $\neg F$.~\qed

}

The theorem may suggest that redundancy is preserved across revisions, that a
redundant formula remains so in spite of further revisions. As a matter of
fact, this is not what the theorem proves. Revising appends a formula to the
sequence. Yet, the theorem extends to that case as well. Still better, it
extends to adding formulae in arbitrary points of the sequence.

\state{between}{theorem}{}{

\begin{theorem}
\statelabel

If
{} $\emptyset [F,S_1,\ldots,S_{i-1},S_i,\ldots S_m] =
{}  \emptyset   [S_1,\ldots,S_{i-1},S_i,\ldots S_m]$
then
{} $\emptyset [F,S_1,\ldots,S_{i-1},S_a,S_i,\ldots S_m] = 
{}  \emptyset   [S_1,\ldots,S_{i-1},S_a,S_i,\ldots S_m]$.

\end{theorem}

}{

\proof Every Q-combination
{} $Q = (B_1 \equiv S_1) \wedge \cdots \wedge (B_{i-1} \equiv S_{i-1}) \wedge
{}      (B_i \equiv S_i) \wedge \cdots \wedge (B_m \equiv S_m)$
of
{} $S_1,\ldots,S_{i-1},S_i,\ldots S_m$
entails either $F$ or $\neg F$ by Theorem~\ref{relevance}.

The Q-combinations
{} $S_1,\ldots,S_{i-1},S_a,S_i,\ldots S_m$
are
{} $Q' = (B_1 \equiv S_1) \wedge \cdots \wedge (B_{i-1} \equiv S_{i-1}) \wedge
{}       (B_a \equiv S_a) \wedge
{}       (B_i \equiv S_i) \wedge \cdots \wedge (B_m \equiv S_m)$.
Each comprises a Q-combination of
{} $S_1,\ldots,S_{i-1},S_i,\ldots S_m$
and $B_a \equiv S_a$. As a result, it entails $Q$. Since every $Q$ entails
either $F$ or $\neg F$, the same does $Q'$. The claim follows by
Theorem~\ref{relevance}.~\qed

}

Future revisions follow the sequence. They are appended to it. The theorem
specializes to this case as follows.

\begin{corollary}
\label{last-following}

If
{} $\emptyset [F,S_1,\ldots,S_m] = \emptyset [S_1,\ldots,S_m]$
then
{} $\emptyset [F,S_1,\ldots,S_m,S_{m+1}] = \emptyset [S_1,\ldots,S_m,S_{m+1}]$.

\end{corollary}

\

Another question is whether redundancy is affected by formulae that precede the
redundant formula. Technically: does the redundancy of $F$ in $[F] \cdot S$
imply its redundancy in $R \cdot [F] \cdot S$?

\state{equivalence-part}{theorem}{}{

\begin{theorem}
\statelabel

If $\emptyset S = \emptyset S'$ then
$\emptyset R \cdot S = \emptyset R \cdot S'$.

\end{theorem}

}{

\proof The premise $\emptyset S = \emptyset S'$ is that $I \leq_{\emptyset S}
J$ coincides with $I \leq_{\emptyset S'} J$ for all models $I$ and $J$. The
claim $\emptyset R \cdot S = \emptyset R \cdot S'$ is that $I \leq_{\emptyset R
\cdot S} J$ coincides with $I \leq_{\emptyset R \cdot S'} J$ for all models $I$
and $J$.

The first condition of the conclusion is $I \leq_{\emptyset R \cdot S} J$.
Lemma~\ref{break} reformulates it as follows.

\[
I \leq_{\emptyset S} J
\mbox{ AND } 
(J \not\leq_{\emptyset S} I \mbox{ OR } I \leq_{\emptyset R} J)
\]

The premise $\emptyset S = \emptyset S'$ is defined as $I \leq_{\emptyset S} J$
being the same as $I \leq_{\emptyset S'} J$. Nothing changes by replacing the
former with the latter.

\[
I \leq_{\emptyset S'} J
\mbox{ AND } 
(J \not\leq_{\emptyset S'} I \mbox{ OR } I \leq_{\emptyset R} J)
\]

Lemma~\ref{break} tells this condition equivalent to $I \leq_{\emptyset R \cdot
S'} J$.

This chain of equivalences proves the first condition $I \leq_{\emptyset R
\cdot S} J$ equivalent to the last $I \leq_{\emptyset R \cdot S'} J$: the
orders $\emptyset R \cdot S$ and $\emptyset R \cdot S'$ coincide.~\qed

}

A consequence of this theorem is that previous revisions do not invalidate
redundancy.

\begin{corollary}
\label{redundant-elongated}

If
{} $\emptyset S \cdot R = S \emptyset \cdot [F] \cdot R$
then
{} $\emptyset Q \cdot S \cdot R = \emptyset Q \cdot S \cdot [F] \cdot R$.

\end{corollary}

\begin{corollary}
\label{redundant-preceding}

If $\emptyset S = \emptyset [F] \cdot S$ then
$\emptyset R \cdot S = \emptyset R \cdot [F] \cdot S$.

\end{corollary}

The converse to Corollary~\ref{redundant-preceding} is generally false, as
shown by the following counterexample.

\state{redundancy-uncut}{counterexample}{}{

\begin{counterexample}
\statelabel

There exists two sequence of lexicographic revisions $S$ and $R$ and a formula
$F$ such that
{} $\emptyset R \cdot S = \emptyset R \cdot [F] \cdot S$
holds and
{} $\emptyset S = \emptyset [F] \cdot S$
does not.

\end{counterexample}

}{

\proof The orders $S$ and $R$ and the formula $F$ are as follows.

\begin{eqnarray*}
S &=& [a] 			\\
F &=& \neg a \vee b		\\
R &=& [a \wedge b]
\end{eqnarray*}

The claim is
{} $\emptyset S \not= \emptyset [F] \cdot S$
and
{} $\emptyset R \cdot S = \emptyset R \cdot [F] \cdot S$.

\begin{itemize}

\item

The first part of the claim
{} $\emptyset S \not= \emptyset [F] \cdot S$
is a consequence of $I \leq_{\emptyset S} J$, $J \leq_{\emptyset S} I$ and $I
\not\leq_{\emptyset F} J$ as proved by
Corollary~\ref{equivalent-made-not-single}. The models $I$ and $J$ are the
following ones.

\begin{eqnarray*}
I &=& \{a, \neg b\}		\\
J &=& \{a, b\}
\end{eqnarray*}

Both models satisfy the only formula $a$ of $S$. Therefore, both $I
\leq_{\emptyset a} J$ and $J \leq_{\emptyset a} I$ hold. When a sequence $S$
comprises a single formula $a$, the definition of lexicographic revision
specializes to
{} $I \leq_{\emptyset a} J \mbox{ and }
{}  ( J \not\leq_{\emptyset a} I \mbox{ or } I \leq_{\emptyset []} J )$,
which is the same as $I \leq_{\emptyset a} J$ since $I \leq_{\emptyset []} J$
holds by definition. As a result, both $I \leq_{\emptyset S} J$ and $J
\leq_{\emptyset S} I$ hold.

Since $I$ falsifies $F$ and $J$ satisfies it, $I \leq_{\emptyset F} J$ does not
hold.

\item

The second part of the claim is
{} $\emptyset R \cdot S = \emptyset R \cdot [F] \cdot S$:
the two orders compare models equally. It can be proved by calculating the
orders between all four models, but most models can be accounted for in groups.

Two models $I$ and $J$ are assumed by contradiction to being sorted differently
by the two orders.

If $I \leq_{\emptyset a} J$ and $J \not\leq_{\emptyset a} I$ both hold, the
definition of an order with $a$ as its last revision specializes from
{} ``$I \leq_{\emptyset a} J$ and
{}   either $J \not\leq_{\emptyset a} I$ or $I \leq_{\emptyset Q} J$''
where $Q$ is sequence of the preceding revisions to
{} ``$\true$ and either $\true$ or $I \leq_{\emptyset Q} J$'',
which is true. As a result, $I$ is less than or equal than $J$ in both orders.
The case where $I \leq_{\emptyset a} J$ and $J \not\leq_{\emptyset a} I$ is
excluded: $I$ satisfies $a$ and $J$ does not.

By symmetry, that $I$ falsifies $a$ and $J$ satifies is excluded as well.

The conclusion is that $I$ and $J$ set $a$ to the same value.

As a result, $I \leq_{\emptyset a} J$ and $J \leq_{\emptyset a} I$ both hold.
Every order $\emptyset Q \cdot [a]$ specializes from
{} ``$I \leq_{\emptyset a} J$ and
{}   either $J \not\leq_{\emptyset a} I$ or $I \leq_{\emptyset Q} J$''
to
{} ``$\true$ and either $\false$ or $I \leq_{\emptyset Q} J$'',
which is the same as $I \leq_{\emptyset Q} J$, where $Q$ is the sequence of the
preceding revisions. Therefore,
{} $\emptyset R \cdot [F] \cdot S$
is the same as
{} $\emptyset R \cdot [F]$
and
{} $\emptyset R \cdot S$
is the same as
{} $\emptyset R$.

The assumption is that
{} $\emptyset R \cdot S$
and
{} $\emptyset R \cdot [F] \cdot S$
compare $I$ and $J$ differently. It implies that $R$ and $R \cdot [F]$ do the
same.

The two models set $a$ to the same value.

If both models set $a$ to $\false$, they both satisfy $F = \neg a \vee b$. As a
result, both $I \leq_{\emptyset F} J$ and $J \leq_{\emptyset F} I$ hold. The
definition of the first order $R \cdot [F]$ specializes from
{} ``$I \leq_{\emptyset F} J$ and
{}   either $J \not\leq_{\emptyset F} I$ or $I \leq_{\emptyset R} J$''
to
{} ``$\true$ and either $\false$ or $I \leq_{\emptyset R} J$'',
which is the same as $I \leq_{\emptyset R} J$, the second order. The assumption
that the order compare $I$ and $J$ differently is contradicted.

The only remaining case is that both models set $a$ to $\true$.

Setting $a$ to true makes $F = \neg a \vee b$ equivalent to $b$ and $a \wedge
b$ equivalent to $b$. This makes $\emptyset R \cdot [F]$ coincide with
$\emptyset [b,b]$ and $\emptyset R$ with $\emptyset [b]$.

The order $\leq_{\emptyset [b]}$ is defined as
{} ``$I \leq_{\emptyset b} J$ and
{}   either $J \not\leq_{\emptyset b} I$ or $I \leq_{\emptyset []} J$'',
which is the same as
{} ``$I \leq_{\emptyset b} J$ and
{}   either $J \not\leq_{\emptyset b} I$ or $\true$''
since $I \leq_{\emptyset []} J$ always holds by definition, which is the same
as
{} ``$I \leq_{\emptyset b} J$ and $\true$'',
or $I \leq_{\emptyset b} J$.

The order $\leq_{\emptyset [b,b]}$ is defined as
{} ``$I \leq_{\emptyset b} J$ and
{}   either $J \not\leq_{\emptyset b} I$ or $I \leq_{\emptyset [b]} J$'',
which is the same as
{} ``$I \leq_{\emptyset b} J$ and
{}   either $J \not\leq_{\emptyset b} I$ or $I \leq_{\emptyset b} J$''
since $I \leq_{\emptyset [b]} J$ is proved above to be equivalent to $I
\leq_{\emptyset b} J$. The first conjunct $I \leq_{\emptyset b} J$ entails the
second disjunct $I \leq_{\emptyset b} J$ and therefore entails the whole
disjunction. Therefore, the order is the same as $I \leq_{\emptyset b} J$,
which is the same as the other order.

The same applies by symmetry to $I$ and $J$ swapped.

The conclusion is that the orders compare these models equally, contradicting
the assumption that they do not.

\end{itemize}~\qed

}

While the converse to Corollary~\ref{redundant-preceding} is generally false,
it is true in a particular case.

\state{redundancy-repeat}{theorem}{}{

\begin{theorem}
\statelabel

If $S_i$ is a formula of the sequence $S$, then $\emptyset [S_i] \cdot S$ and
$\emptyset [\neg S_i] \cdot S$ coincide with $\emptyset S$.

\end{theorem}

}{

\proof All conjunctions
{} $Q = (B_1 \equiv S_1) \wedge \cdots \wedge (B_m \equiv S_m)$
contain $B_i \equiv S_i$, where $B_i$ is either $\true$ or $\false$. If $B_i$
is $\true$ then $B_i \equiv S_i$ is equivalent to $S_i$, otherwise it is
equivalent to $\neg S_i$. Either way, $Q$ contains either $S_i$ or $\neg S_i$
and therefore entails it. Theorem~\ref{relevance} proves that $\emptyset [S_i]
\cdot S$ coincide with $\emptyset S$ since every $Q$ entails either $S_i$ or
$\neg S_i$. The theorem also proves that $\emptyset [\neg S_i] \cdot S$
coincide with $\emptyset S$ since every $Q$ entails either $\neg S_i$ or $\neg
\neg S_i$.~\qed

}

\section{Short Horn sequences of lexicographic revisions}
\label{horn}

Checking redundancy in a sequence of Horn formulae from the flat doxastic state
is \conp-hard~\cite{libe-24-a}. It is as hard as in the general case, but only
for arbitrarily long sequences. Bounding the number of revision to two makes it
polynomial.

\subsection{Binary Horn sequences}
\label{horn-two}

A polynomial-time algorithm establishes the redundancy of the first revision in
a sequence of two Horn revisions from the flat doxastic state. It tells whether
$\emptyset [S_1,S_2]$ coincides with $\emptyset [S_2]$, whether forgetting
$S_1$ causes no information loss.

Redundancy holds in four possible cases thanks to Theorem~\ref{relevance-two}:
$S_1$ is either inconsistent, tautological, equivalent to $S_2$ or equivalent
to $\neg S_2$. Each case can be established in polynomial time when both $S_1$
and $S_2$ are Horn. Inconsistency and validity are checked in polynomial time
by unit propagation~\cite{dowl-gall-84,bier-etal-21}. Equivalence is the same
as double entailment: $S_1 \models S_2$ and $S_2 \models S_1$; entailment is
the same as the entailment of every clause of the entailed formula, and $S_i
\models C$ is equivalent to the unsatisfiability of $S_i \wedge \neg C$, which
is a Horn formula since $\neg C$ is a conjunction of literals. Unsatisfiability
of a Horn formula is polynomial-time.

The only remaining case is the equivalence between $S_1$ and $\neg S_2$,
between a Horn formula and a negated Horn formula. Equivalence is again the
same as two entailments. The first $S_1 \models \neg S_2$ is easy since it is
same as the inconsistency of $S_1 \wedge S_2$, which is Horn. The second $\neg
S_2 \models S_1$ is not, since $\neg S_2 \wedge \neg S_1$ is not a Horn
formula.

A polynomial-time algorithm for verifying $S_1 \equiv \neg S_2$ is shown. It
works because equivalence rarely holds, and the few cases where it does are
easy to check.

The following lemma gives a sufficient condition to non-equivalence.

\state{positive}{lemma}{}{

\begin{lemma}
\statelabel

If a formula entails a positive clause and none of its literals, it is not
equivalent to any Horn formula.

\end{lemma}

}{

\proof Contrary to the claim, a Horn formula $F$ is assumed to entail a
positive clause $x_1 \vee \cdots \vee x_k$ but none of its literals:

\begin{eqnarray*}
F &\models& x_1 \vee \cdots \vee x_k \\
F &\not\models& x_1 \\
\vdots \\
F &\not\models& x_k
\end{eqnarray*}

Each condition $F \not\models x_i$ is equivalent to the satisfiability of $F
\wedge \neg x_i$: for every $i$, a model $I_i$ satisfies $F$ but not $x_i$.

Horn formulae have a property about the combination of every two of its models.
The intersection of two models is the model that sets to true exactly the
variables that they both set to true. For example, the intersection between
{} $\{a,      b, \neg c\}$ and
{} $\{a, \neg b,      c\}$ is the model
{} $\{a, \neg b, \neg c\}$.

Since $F$ is satisfied by the models $I_1,\ldots,I_k$, it is also satisfied by
their intersection. Each $I_i$ sets $x_i$ to false. Therefore, their
intersection sets all variables $x_i$ to false. It therefore falsifies $x_1
\vee \cdots \vee x_k$, contrary to the assumption that $F$ entails this
clause.~\qed

}

The rest of the proof requires deleting a variable from a formula by removing
the clauses containing an instance of the variable and by removing the negation
of the variable from the other clauses. The result is denoted $F\%x$.

\[
F\%x =
\{f' \mid \exists f \in F ~.~ x \not\in f ,~ f' = f \backslash \{\neg x\}\}
\]

\draft

clause contains x   ===>  remove clause

clause contains -x  ===>  remove -x

\enddraft

This removal is inessential if the formula entails a variable or is entailed by
its negation.

\state{remove-literal}{lemma}{}{

\begin{lemma}
\statelabel

If $F \models x$, then
{} $F \equiv x \wedge (F\%x)$.

If $\neg x \models F$, then
{} $F \equiv \neg x \vee (F\%x)$.

\end{lemma}

}{

\proof The two claims are proved in turn.

\begin{description}

\item[$F \models x$:]\

This entailment implies the equivalence of $F$ and $F \wedge x$. In this
formula:

\begin{enumerate}

\item all clauses of $F$ that contain $x$ are redundant as entailed by $x$ and
can therefore be removed;

\item $\neg x$ is false and can be therefore removed from all clauses of $F$
containing it.

\end{enumerate}

The result of this simplification is $F\%x$. As a result, $F$ is equivalent to
$x \wedge (F\%x)$.

\item[$\neg x \models F$:]\

Since $\neg x$ entails the conjunction of clauses $F$, it entails all of them.
This is only possible if all clauses contain $\neg x$. The clauses also
containing $x$ are tautological and can be removed. The remaining clauses still
contain $\neg x$, which can be factored out. The result is $\neg x \vee F'$
where $F'$ is $F$ after removing the clauses containing $x$ and removing the
clauses containing $\neg x$. The result of these removals is exactly $F\%x$.
This proves the equivalence of $F$ with $\neg x \vee F'$.

\end{description}~\qed

}

This lemma gives a way to simplify equivalence when one of the two formulae
entails a variable or is entailed by its negation.

It works even if $F$ is inconsistent. An example is $F=\{x,\neg x\}$. It
entails $x$. The removal of $x$ from $F$ deletes its first clause because it
contains $x$ and removes $\neg x$ from its second clause, leaving it empty. The
result is $F\%x = \{\emptyset\}$. An empty clause is unsatisfiable, making $x
\wedge (F\%x)$ inconsistent and therefore equivalent to $F$.

\state{simplify}{lemma}{}{

\begin{lemma}
\statelabel

\

\begin{itemize}

\item
If $F_2 \models x$ holds, then 
$\neg F_2 \equiv F_1$ holds if and only if
both $\neg x \models F_1$ and $\neg (F_2\%x) \equiv F_1\%x$ hold;

\item If $\neg x \models F_2$ holds, then
$\neg F_2 \equiv F_1$ holds if and only if
both $F_1 \models x$ and $\neg (F_2\%x) \equiv F_1\%x$ hold.

\end{itemize}

\end{lemma}

}{

\proof Only the first statement is proved. The second is symmetric.


The first statement is proved in each direction. The first part of the proof
assumes
{} $F_2 \models x$
and
{} $\neg F_2 \equiv F_1$
and proves
{} $\neg (F_2\%x) \equiv F_1\%x$.
The second assumes 
{} $F_2 \models x$,
{} $\neg x \models F_1$
and
{} $\neg (F_2\%x) \equiv F_1\%x$,
and proves
{} $\neg F_2 \equiv F_1$.




The first direction assumes
{} $F_2 \models x$
and
{} $\neg F_2 \equiv F_1$.

The first assumption $F_2 \models x$ implies $\neg x \models \neg F_2$ by modus
tollens. This is the same as $\neg x \models F_1$ since $\neg F_2$ is
equivalent to $F_1$ by assumption. This is the first part of the claim.

{} Lemma~\ref{remove-literal}
proves $F_1 \equiv \neg x \vee (F_1\%x)$ from $\neg x \models F_1$. Negating
both sides of the equivalence gives $\neg F_1 \equiv x \wedge \neg (F_1\%x)$.
{} Lemma~\ref{remove-literal}
also proves $F_2 \equiv x \wedge (F_2\%x)$ from $F_2 \models x$. These two
equivalences are linked by $F_2 \equiv \neg F_1$, a consequence of the
assumption $\neg F_2 \equiv F_1$. They imply
{} $x \wedge \neg (F_1\%x) \equiv x \wedge (F_2\%x)$.
Since neither $F_1\%x$ nor $F_2\%x$ contain $x$, this is the same as
{} $\neg (F_1\%x) \equiv (F_2\%x)$.
Negating both sides of the equivalence gives the second claim:
{} $(F_1\%x) \equiv \neg (F_2\%x)$.



The second direction assumes
{} $F_2 \models x$,
{} $\neg x \models F_1$
and
{} $\neg (F_2\%x) \equiv (F_1\%x)$.

Reversing both sides of the equivalence gives
{} $\neg (F_1\%x) \equiv (F_2\%x)$.
The first assumption
{} $F_2 \models x$
and
{} Lemma~\ref{remove-literal}
imply
{} $F_2 \equiv x \wedge (F_2\%x)$.
The second assumption
{} $\neg x \models F_1$
and
{} Lemma~\ref{remove-literal}
imply
{} $F_1 \equiv \neg x \vee (F_1\%x)$.
Reversing both sides of this equivalence gives
{} $\neg F_1 \equiv x \wedge \neg (F_1\%x)$.
Since
{} $\neg (F_1\%x) \equiv (F_2\%x)$,
this is the same as
{} $\neg F_1 \equiv x \wedge (F_2\%x)$.
This equivalence and
{} $F_2 \equiv x \wedge (F_2\%x)$
imply
{} $\neg F_1 \equiv F_2$,
the same as
{} $F_1 \equiv \neg F_2$,
the claim.~\qed






}

Since $\neg F_2 \equiv F_1$ is the same as $F_2 \equiv \neg F_1$, the lemma
also applies in the other direction. Equivalence simplifies if either formula
entails a variable or is entailed by its negation.

\begin{algorithm}
\label{algorithm-simplify}

\

\noindent
while either $F_1$ or $F_2$ change:

\begin{enumerate}

\item if $F_2 \models x$:

\begin{enumerate}

\item if $\neg x \not\models F_1$ then return $\false$

\item $F_1 = F_1\%x$
\item $F_2 = F_2\%x$

\end{enumerate}

\item if $\neg x \models F_2$:

\begin{enumerate}

\item if $F_1 \not\models x$ then return $\false$

\item $F_1 = F_1\%x$
\item $F_2 = F_2\%x$

\end{enumerate}

\item if $F_1 \models x$:

\begin{enumerate}

\item if $\neg x \not\models F_2$ then return $\false$

\item $F_1 = F_1\%x$
\item $F_2 = F_2\%x$

\end{enumerate}

\item if $\neg x \models F_1$:

\begin{enumerate}

\item if $F_2 \not\models x$ then return $\false$

\item $F_1 = F_1\%x$
\item $F_2 = F_2\%x$

\end{enumerate}

\end{enumerate}

\noindent
if $F_1 = \emptyset$ then return $F_2 \models \false$

\noindent
if $F_1 \models \false$ then return $\models F_2$

\noindent
if $F_2 = \emptyset$ then return $F_1 \models \false$

\noindent
if $F_2 \models \false$ then return $\models F_1$

\item return $\false$

\end{algorithm}

The loop simplifies the two formulae as much as possible according to
Lemma~\ref{simplify}. It leaves them very simple to check: one is inconsistent
and the other valid.

\state{algorithm-simplify-empty}{lemma}{}{

\begin{lemma}
\statelabel

Algorithm~\ref{algorithm-simplify} checks whether $F_1 \equiv \neg F_2$
if both $F_1$ and $F_2$ are Horn formulae.

\end{lemma}

}{

\proof The algorithm first simplifies $F_1$ and $F_2$ as much as possible.
Lemma~\ref{simplify} proves that this simplification is correct: if $\false$ is
returned then the two formulae are not equivalent; otherwise, the condition
$F_1 \equiv \neg F_2$ is unchanged by the removal of $x$ from both formulae.

The four steps following the loop are also correct: an empty formula is
tautological and therefore equivalent only to the negation of a contradiction;
a contradictory formula is only equivalent to a tautology.

The final step of the algorithm is only reached if neither formula is either
empty or contradictory. Because of the previous simplification, they do not
contain any variable they entail or are entailed by its negation.

Since $F_2$ is not empty, it contains a clause. That clause is not empty since
an empty clause is a contradiction and $F_2$ is not contradictory.

A non-empty clause of $F_2$ does not comprise a single positive literal since
$F_2$ would otherwise entail that single positive literal and the loop would
have removed the clause. Since this clause is not empty and does not comprise a
single positive literal, it comprises either a single negative literal or two
literals at least. In the second case, at least one of these literals is
negative since Horn clauses do not contain more than one positive literal. In
both cases, the clause contains at least a negative literal.

If this negative literal were the same for all clauses, it would entail every
clause of $F_2$ and therefore the whole $F_2$. The loop would have removed it.

Consequently, $F_2$ does not contain a single clause. Otherwise, the negative
literal of the clause would be the same for all clauses of $F_2$.

The conclusion is that $F_2$ contains at least two clauses, each containing a
negative literal not contained in all other clauses. As a result, $F_2$ formula
can be rewritten as follows.

\[
(\neg x \vee c_1) \wedge
(\neg y \vee c_2) \wedge
(\neg z \vee c_3) \wedge
\ldots
\]

The negation of this formula is as follows.

\begin{eqnarray*}
\lefteqn{
\neg(
(\neg x \vee c_1) \wedge
(\neg y \vee c_2) \wedge
(\neg z \vee c_3) \wedge
\ldots
)}
\\
&\equiv&
(x \wedge \neg c_1) \vee
(y \wedge \neg c_2) \vee
(z \wedge \neg c_3) \vee
\ldots
)
\\
&\equiv&
(x \vee y \vee z \vee \cdots) \wedge \ldots
\end{eqnarray*}

The positive clause comes from reversing the sign of each negative literal of
each clause. As proved above, $F_2$ contains at least two clauses, each
containing a negative literal at least, and these literals cannot be all the
same. As a result, the negation of $F_2$ contains a positive clause comprising
at least two literals.

The negation of $F_2$ is assumed equivalent to $F_1$.

Since $F_1$ entails the negation of $F_2$, it also entails its clause that
comprises two positive literals or more. Since $F_1$ is Horn, it entails at
least one of these variables by Lemma~\ref{positive}. Since it is satisfiable,
it contains that variable. This is a contradiction since $F_1$ would have been
removed that variable in the loop.

The only assumption is that $\neg F_2$ is equivalent to $F_1$. Therefore, it is
not. Returning $\false$ is correct.~\qed

}

The algorithm is not only correct, it is also polynomial-time.

\state{algorithm-simplify-polynomial}{lemma}{}{

\begin{lemma}
\statelabel

Algorithm~\ref{algorithm-simplify} runs in polynomial time.

\end{lemma}

}{

\proof The loop is executed until the formulae stop changing. Therefore, its
iterations are no more than the size of the formulae.

Each step is linear: $F \models x$ is equivalent to the inconsistency of the
Horn formula $F \cup \{\neg x\}$, which can be checked in linear time; $\neg x
\models F$ is the same as $\neg x$ being contained in all clauses of $F$.

The following steps are linear as well, since they only require checking the
inconsistency of Horn formulae.~\qed

}

The algorithm is polynomial-time and correctly identifies Horn formulae that
are one equivalent to the negation of the other. This was the only case left of
the four that make redundancy.

\state{horn-two-polynomial}{theorem}{}{

\begin{theorem}
\statelabel

Checking the redundancy of the first of a sequence of two Horn revisions from
the flat doxastic state takes polynomial time.

\end{theorem}

}{

\proof Theorem~\ref{relevance-two} proves that $\emptyset [S_1,S_2] = \emptyset
[S_2]$ holds in exactly four cases. They can be checked in polynomial time.

\begin{description}

\item[$S_1$ inconsistent:] polynomial-time since $F_1$ is Horn;

\item[$S_1$ tautological:] only possible if all its clauses are tautological;

\item[$S_1 \equiv S_2$:] is the same as $S_1 \models S_2$ and $S_2 \models
S_1$; the first is the same as $S_1 \models f$ for each $f \in S_2$, which is
the same as the inconsistency of the Horn formula $S_1 \cup \{\neg l \mid l \in
f\}$; the same for $S_2 \models S_1$;

\item[$S_1 \equiv \neg S_2$:] Algorithm~\ref{algorithm-simplify} checks this
condition as proved by Lemma~\ref{algorithm-simplify-empty} and takes
polynomial time as proved by Lemma~\ref{algorithm-simplify-polynomial}.

\end{description}~\qed

}

\subsection{Ternary Horn sequences}
\label{horn-three}

The redundancy problem is coNP-complete on revision sequences of arbitrary
length~\cite{libe-24-a} and is polynomial on sequences of two as proved in the
previous section. A question is whether or not there is a particular sequence
length at which the complexity jump occurs. For example, the redundancy problem
in a triple is whether $\emptyset [S_1,S_2,S_3]$ the same as $\emptyset
[S_2,S_3]$. Even if the sequences are only elongated by a single formula, it is
unclear that the methods used for proving polynomiality extend to cover this
case.

The polynomial-time algorithm for sequences of two formulae hinges around the
reformulation given by Theorem~\ref{relevance-equivalence}. For three formulae,
it is the equivalence with any of the following formulae:
\begin{eqnarray*}
\false						\\
S_2 \wedge S_3					\\
S_2 \wedge \neg S_3				\\
\neg S_2 \wedge S_3				\\
\neg S_2 \wedge \neg S_3			\\
S_2 \wedge S_3 \vee S_2 \wedge \neg S_3		\\
S_2 \wedge S_3 \vee \neg S_2 \wedge \neg S_3	\\
S_2 \wedge S_3 \vee S_2 \wedge \neg S_3		\\
S_2 \wedge S_3 \vee \neg S_2 \wedge \neg S_3	\\
\vdots						\\
S_2 \wedge S_3 \vee S_2 \wedge \neg S_3 \vee \neg S_2 \wedge \neg S_3 \\
\vdots						\\
\true
\end{eqnarray*}

Some of these formulae are equivalent to either $S_2$ or $S_3$ or their
negations; checking them is easy. The others are not. An example is $S_2 \wedge
\neg S_3$. This formula may not be Horn.

The same problem is overcome for orders of two formulae by
Algorithm~\ref{algorithm-simplify}. It rules out the cases where a Horn formula
is not equivalent to the negation of another Horn formula. Whether $S_1$ is
equivalent to $\neg S_2$ is established this way.

This is not enough for three formulae or more. Equivalence between $S_1$ and
$S_2 \wedge \neg S_3$ does not require $S_1$ to be equivalent to a Horn formula
or the negation of a Horn formula. Even if $S_2$ is Horn and $\neg S_3$ is not
equivalent to a Horn formula, their conjunction may still be. An example
follows.

\[
\{-x \vee a\} \cup \{x, a \vee b\}
\]

The first part of the union is a Horn formula and the second is not equivalent
to any. Yet, the whole set entails $a$, which makes $a \vee b$ redundant. It is
equivalent to $\{x, a\}$, which is Horn.

This proves that ruling out any formula to be equivalent to the negation of
another Horn formula is not enough. Such an equivalence is the premise of
Lemma~\ref{simplify}, the engine of Algorithm~\ref{algorithm-simplify}.

\section{Conclusions}
\label{conclusions}

\draft
{\bf SUMMARY OF ARTICLE}
\enddraft

Forgetting may not cause information loss because the same information may be
implied by what is remembered. The question applies for example to privacy,
where concealing a specific source of information may not be enough to hide
sensitive information~\cite{gonc-etal-17}. It also apply to many other
applications of forgetting, such as
{} psychology~\cite{myer-04}
(does forgetting an uncanny experience really cancels its consequences),
{} reasoning simplification~\cite{delg-wang-15,erde-ferr-07,wang-etal-05},
{} knowledge clarification~\cite{delg-17}
{} and searching reduction~\cite{smyt-kean-95}
(does forgetting erase essential rather than marginal information?).

In the context of iterated belief revision, information is acquired by a
sequence of revisions, each providing a piece of information in terms of a
logical formula. The question is whether a revision can be removed from the
sequence while maintaining the final result---the current state of beliefs.

The computational study of the problem~\cite{libe-24-a} left two open
questions.

The first is the complexity of heterogeneous sequences of revisions, which
simplifies to \conp-complete for homogeneous sequences of lexicographic
revisions. This result is enhanced by proving heterogeneous sequence \dpi-hard.
Membership to \conp\  would imply a collapse in the polynomial hierarchy to its
second level, something currently considered unlikely in the field of
computational complexity~\cite{aror-bara-09}.

The second open question is the complexity of short sequences of Horn
lexicographic revisions. The problem is \conp-complete even for sequences of
two arbitrary revisions, and \conp-hard for unbound sequences of Horn
revisions. A polynomial algorithm is shown for the case of two Horn revisions.
A side technical result is a polynomial algorithm for checking the equivalence
of a Horn formula with the negation of a Horn formula.

\draft
{\bf FUTURE DIRECTIONS} 
\enddraft

The questions are not closed.

Complexity is polynomial for two lexicographic Horn revisions and \conp-hard
for an unbounded number of them, but the boundary could be at three or could be
at a linear number of revisions. Three revisions look hard at a first
inspection, but no proof was found.

The same goes for heterogenous sequences of revisions. Completeness proved for
\dpi\  could be raised to \D2 or for one of the many classes in between.
Membership to \D2 could be lowered up to \dpi.

Other kinds of revisions~\cite{cant-97,rott-06, arav-etal-18, gura-kodi-18,
souz-etal-19, benf-etal-93, kuts-19, andr-etal-02, saue-etal-22, doms-etal-11,
boot-chan-17, boot-chan-20} may have a different complexity, especially if they
differ on how they represent the doxastic state~\cite{ryan-91, will-92,
dixo-93, will-95, dixo-wobc-93}.


\appendix

\bibliographystyle{alpha}
\newcommand{\etalchar}[1]{$^{#1}$}

\end{document}